\documentclass[journal,letterpaper,twocolumn,final,10pt]{IEEEtran}

\usepackage{amsthm}
\usepackage{amsmath}
\usepackage{amsfonts}
\usepackage{dsfont}
\usepackage{mathrsfs}
\usepackage{pifont}
\usepackage{amssymb}
\usepackage{verbatim}
\usepackage{upgreek}
\usepackage{color}
\usepackage[final]{graphicx}
\usepackage{epstopdf}
\usepackage{algorithmic}
\usepackage{algorithm}
\usepackage{epsfig}
\usepackage{eucal}
\usepackage{cite}
\usepackage{subfigure}
\usepackage{balance}
\usepackage{setspace}
\usepackage{siunitx}
\usepackage{gensymb}

\makeatletter
\def\maketag@@@#1{\hbox{\m@th\normalfont\normalsize#1}}
\newcommand{\Rmnum}[1]{\expandafter\@slowromancap\romannumeral #1@}
\makeatother

\newcommand{\bGamma}{\mbox{\boldmath{$\Gamma$}}}

\newcommand{\bbm}{\begin{bmatrix}}
\newcommand{\ebm}{\end{bmatrix}}

\newcommand{\bit}{\begin{itemize}}
\newcommand{\eit}{\end{itemize}}

\newcommand{\ben}{\begin{enumerate}}
\newcommand{\een}{\end{enumerate}}

\newcommand{\bdesc}{\begin{description}}
\newcommand{\edesc}{\end{description}}

\newcommand{\bea}{\begin{array}}
\newcommand{\eea}{\end{array}}

\newcommand{\tr}{\mbox{\rm Tr}\, }

\newcommand{\beqa}{\begin{eqnarray}}
\newcommand{\eeqa}{\end{eqnarray}}

\newcommand{\ds}{\displaystyle}

\newcommand{\Comment}[1]{}

\def\C{{\mathds C}}

\def\ccL{\mbox{$\mathfrak L$}}
\def\cA{\mbox{$\mathcal A$}}

\def\cC{\mbox{$\CMcal C$}}

\def\cL{\mbox{$\mathcal L$}}
\def\cN{\mbox{$\CMcal N$}}

\def\cP{\mbox{$\mathcal P$}}

\newcommand{\be}{\begin{equation}}

\newcommand{\ee}{\end{equation}}

\newcommand{\bzero}{{\mbox{\boldmath $0$}}}

\newcommand{\bm}{{\mbox{\boldmath $m$}}}
\newcommand{\bp}{\mbox{\boldmath $p$}}
\newcommand{\bor}{{\mbox{\boldmath $r$}}}

\newcommand{\bv}{{\mbox{\boldmath $v$}}}
\newcommand{\bx}{{\mbox{\boldmath $x$}}}

\newcommand{\bz}{{\mbox{\boldmath $z$}}}

\newcommand{\bA}{{\mbox{\boldmath $A$}}}
\newcommand{\bB}{{\mbox{\boldmath $B$}}}

\newcommand{\bD}{{\mbox{\boldmath $D$}}}

\newcommand{\bI}{{\mbox{\boldmath $I$}}}

\newcommand{\bM}{{\mbox{\boldmath $M$}}}

\newcommand{\bQ}{{\mbox{\boldmath $Q$}}}
\newcommand{\bR}{{\mbox{\boldmath $R$}}}
\newcommand{\bS}{{\mbox{\boldmath $S$}}}
\newcommand{\bT}{{\mbox{\boldmath $T$}}}
\newcommand{\bU}{{\mbox{\boldmath $U$}}}
\newcommand{\bV}{{\mbox{\boldmath $V$}}}
\newcommand{\bX}{{\mbox{\boldmath $X$}}}

\newcommand{\bZ}{{\mbox{\boldmath $Z$}}}

\newcommand{\diag}{\mbox{diag}\, }

\newcommand{\bLambda}{{\mbox{\boldmath $\Lambda$}}}

\newcommand{\bSigma}{{\mbox{\boldmath $\Sigma$}}}

\newcommand{\dmax}{\begin{displaystyle}\max\end{displaystyle}}
\newcommand{\dmin}{\begin{displaystyle}\min\end{displaystyle}}

\newcommand{\test}{\mbox{$
\begin{array}{c}
\stackrel{ \stackrel{\textstyle H_{1,i}}{\textstyle >} }{
\stackrel{\textstyle <}{\textstyle H_0} }
\end{array}
$}}

\DeclareMathOperator*{\argmax}{arg\,max}

\title{Innovative Cognitive Approaches for Joint Radar Clutter Classification and Multiple Target Detection
in Heterogeneous Environments}

\vspace{0.1cm}

\author{Linjie Yan, Sudan Han, Chengpeng Hao, \IEEEmembership{Senior Member, IEEE},
Danilo Orlando, \IEEEmembership{Senior Member, IEEE}, and Giuseppe Ricci, \IEEEmembership{Senior Member, IEEE}

\thanks{Linjie Yan and Chengpeng Hao are with the Institute of Acoustics, Chinese Academy of Sciences, Beijing
100864, China, E-mail: {\tt yanlinjie16@163.com; haochengp@mail.ioa.ac.cn}. \emph{(Corresponding author: Chengpeng Hao.)}}
\thanks{Sudan Han is with  the  National  Innovation  Institute  of  Defense  Technology, Beijing, China E-mail: xiaoxiaosu0626@163.com.}
\thanks{Danilo Orlando is with Universit\`a degli Studi ``Niccol\`o Cusano'',
via Don Carlo Gnocchi 3, 00166 Roma, Italy. E-mail: {\tt danilo.orlando@unicusano.it}.}
\thanks{Giuseppe Ricci is with the Dipartimento di Ingegneria dell'Innovazione,
        Universit\`a del Salento, Via Monteroni, 73100 Lecce, Italy.
        E-mail: {\tt giuseppe.ricci@unisalento.it}.}
}

\begin{document}

\maketitle

\begin{abstract}
The joint adaptive detection of multiple point-like targets in scenarios characterized by
different clutter types is still an open problem in the radar community. In this paper, we
provide a solution to this problem by devising detection architectures capable of classifying
the range bins according to their clutter properties and detecting possible multiple targets
whose positions and number are unknown. Remarkably, the information provided by the proposed architectures
makes the system aware of the surrounding environment and can be exploited to
enhance the entire detection and estimation performance of the system.
At the design stage, we assume three different signal models and apply the latent variable model
in conjunction with estimation procedures based upon the expectation-maximization
algorithm. In addition, for some models, the maximization step cannot be computed in closed-form
(at least to the best of authors' knowledge)
and, hence, suitable approximations are pursued, whereas, in other cases, the maximization
is exact. The performance of the proposed architectures is assessed over synthetic data
and shows that they can be effective in heterogeneous scenarios providing an initial
snapshot of the radar operating scenario.
\end{abstract}


\section{Introduction}
\label{Sec:Introduction}
The design of adaptive detectors in heterogeneous environments continues to attract the
attention of the radar community. As a matter of fact, it represents a compelling task
and several interesting solutions have been proposed over the years
\cite[and references therein]{COLUCCIA2022108401,clutterClust,Recursive,9145700,381910,1008976,hao2021advances,gini1,gini2}.
The complexity of such solutions and of the considered scenarios is {\em somehow directly proportional}
to the advances in electronic technology and digitalization allowing for
miniaturized high performance sensors and processing boards.

There exist different models to address heterogeneity in radar data that generalize the well-known
homogeneous environment. This model assumes that the range Cell Under Test (CUT) as well as the secondary data,
selected through the (leading and lagging) reference window, share the same statistical characterization
of the interference \cite[and references therein]{Richards,kelly1986adaptive,
robey1992cfar,BOR-Morgan,8809199,Liu2022}.\footnote{In what follows, we use the term interference to denote
the overall disturbance affecting radar data that, generally speaking,
is the sum of three components: thermal noise, clutter, and possible intentional interference.
When we use the term clutter to denote the interference, it is understood that the clutter component
is the most significant.}
The idea is that the interference in the CUT can be equalized by exploiting the estimates of its distribution
parameters obtained through the secondary data. The most important aspect of this approach is that
the corresponding decision rules possess the Constant False Alarm Rate (CFAR) property which allows
radar engineers to set the detection threshold regardless the interference spectral properties.
When the reference window includes range bins characterized by different statistical properties (e.g., different
reflectivity coefficients, different grazing angles, clutter edges, unwanted intentional/unintentional targets, etc.),
the quality of the estimates based upon secondary data
might considerably deplete leading to two main effects \cite{guerci2010cognitive,guerci2014space}:
a detection performance degradation, due to
the attenuation of target components, and an increase of the false alarm rate, due to
undernulled clutter.

An alternative to the homogeneous environment that still allows for mathematical
tractability is represented by the so-called {\em partially-homogeneous environment}, where
interference in the secondary data shares the same structure of the covariance
matrix as in the CUT but a different power level \cite{kraut1999cfar}. It is important to notice that
even though in this case CUT and secondary data are not homogeneous, the statistical characterization of
the training samples is assumed to be the invariant over the range.
In scenarios of practical value, this behavior is not always valid
as corroborated by different experimental measurements \cite{Ward,Ward1990,Farina_Gini_Greco_Verrazzani,
Greco_Gini_Rangaswamy,Wardbook}.
Such results indicate that an effective model for clutter in high-resolution radars or at low grazing angles
is the so-called compound-Gaussian \cite{Compound1,Compound2} formed by the product
of a speckle and a texture component. For short time intervals, the texture component
can be approximated as a constant leading to a {\em fully-heterogeneous environment} where
each range bin is characterized by its own power level \cite{9145700,COLUCCIA2022108401}.

Another approach to deal with heterogeneous interference consists in
the classification of the region under surveillance in terms of clutter properties. Then, only the range bins
belonging to these homogeneous regions come into play to estimate the related distribution parameters.
Thus, for each region a set of estimates is available to make adaptive the decision schemes used
to detect targets in that region.
In this respect, the knowledge-aided paradigm
is an effective means to guide the system towards reliable clutter parameter estimates
\cite{1593334,4014432}. It consists
in accounting for all the available {\em a priori} information about the region of interest to
exclude inhomogeneities from the computation of the Sample Covariance Matrix (SCM) \cite{guerci2010cognitive}.
More recently, classification strategies based upon the latent variable model \cite{murphy2012machine}
and the Expectation-Maximization (EM) algorithm \cite{Dempster77} have been proposed in \cite{clutterClust}
for different models of the disturbance covariance matrix. Such methods partition the region of interest
into homogeneous clusters of (not necessarily contiguous) range bins according to their covariance matrix.
The same kind of clustering but with the constraint of regions formed by contiguous range bins
is developed in \cite{XU2021108127,9707864} where a sliding window moves over the entire radar window and for each position
a test on the presence of a clutter edge is performed. A fusion strategy is also provided to merge
the returned estimates of the clutter edge positions.

However, it is important to observe that modern radar systems can operate in target-rich scenarios where
the risk of incorporating target components into the covariance matrix estimate becomes high.
More importantly, such an estimate is commonly used to whiten data and, hence, the presence of target components
would dramatically attenuate the energy of target echoes with a reduction of the receiver sensitivity.
A tangible example of this problem is provided by the so-called Adaptive Matched Filter with De-emphasis
(AMFD) with the de-emphasis parameter equal to $1$ \cite{751541}
that is equivalent to the Adaptive Matched Filter (AMF) derived in \cite{robey1992cfar} where the SCM
over the secondary data is updated with the CUT. In fact, the AMFD returns poor detection performance with respect
to the AMF and Kelly's detector \cite{BOR-Morgan}.
All these issues point out the need of developing classification schemes not only capable of labeling
the range bins according to the statistical properties of the clutter but also of
identifying the range bins containing structured signals in order to somehow exclude them
from clutter properties estimation. Remarkably, this information can be used to build up
architectures that, unlike classical detectors, jointly process the entire radar window
to detect multiple point-like targets within each clutter region (and, hence,
taking advantage of the entire energy present in the collected data).

With the above remarks in mind, in this paper, we devise radar architectures that jointly classify the clutter returns
in terms of their covariance matrix (by partitioning
the region of interest on the basis of clutter properties) and detect possible multiple point-like targets. It is important
to highlight that the design of this kind of architectures represents the main technical
novelty of this paper. As a matter of fact, existing clutter covariance classifiers do not account
for the possible inhomogeneity due to structured signals backscattered from possible targets
\cite[and references therein]{clutterClust,XU2021108127,9707864}, whereas
detectors of multiple point-like targets assume that clutter returns share the same
covariance matrix \cite[and references therein]{1605248,1275664,4721041,9210231,9456070}.
At the design stage, we consider three target-plus-noise hypotheses that account for different target models
comprising deterministic and fluctuating responses. For each hypothesis, we conceive estimation
procedures exploiting hidden random variables representative of different situations
that can occur at range bin level. To be more definite, the bin under consideration might contain
clutter only characterized
by a specific class of distribution parameters (that identify a clutter region)
or, in addition, target components. Two important remarks are in order. First,
the classification of a range bin as occupied by a target is tantamount to detecting that target;
second, notice that we do not establish any assumption about the position and the number of targets.
In this framework, the application of the Maximum Likelihood Approach (MLA) to come up with
the estimates of the unknown parameters leads to intractable mathematics. For this reason,
we resort to the EM algorithm \cite{Dempster77} (or heuristic
modification of it) in conjunction with cyclic optimization procedures \cite{Stoica_alternating}.
The estimates returned by these procedures are used to build up two decision schemes
based upon the Likelihood Ratio Test (LRT).
The numerical examples are obtained over synthetic data and highlight that the proposed
architectures represent an effective means to detect multiple point-like targets
in heterogeneous scenarios returning reliable detection and classification performance.
More importantly, detection and classification information provided by the proposed
algorithms can be suitably exploited by {\em ad hoc} (possibly conventional)
decision schemes and estimation procedures to enhance the entire performance of the system.

The remainder of the paper is organized as follows. The next section contains the problem formulation along
with useful preliminary definitions, whereas
Section \ref{Sec:Architecture_Designs} is devoted to the design
of the classification architectures. The illustrative examples and the discussion
about the detection and classification performance are provided in Section \ref{Sec:performance}.
Finally, in Section \ref{Sec:conclusions}, we draw the conclusions
and lay down possible future research lines.

\subsection{Notation}
In the sequel, vectors and matrices are denoted by boldface lower-case and upper-case letters, respectively.
Symbols $\det(\cdot)$, $\tr(\cdot)$, $\|\cdot\|$,
and $(\cdot)^\dag$ denote the determinant, trace, Euclidean norm,
and conjugate transpose, respectively.
As to numerical sets,
$\C$ is the set of
complex numbers, and $\C^{N\times M}$ is the Euclidean space of $(N\times M)$-dimensional
complex matrices (or vectors if $M=1$).
The modulus of $x\in\C$ is denoted by $|x|$.
$\bI$ and $\bzero$ stand for the identity matrix and the null vector or matrix of proper size.
Given $a_1, \ldots, a_N \in\C$, $\diag(a_1, \ldots, a_N)\in\C^{N\times N}$ indicates
the diagonal matrix whose $i$th diagonal element is $a_i$.
The acronyms PDF and PMF stand for Probability Density Function and Probability Mass Function, respectively, whereas
the conditional PDF of a random variable $x$ given
another random variable $y$ is denoted by $f(x|y)$.
Finally, we write $\bx\sim\cC\cN_N(\bm, \bM)$ if $\bx$ is a
complex circular $N$-dimensional normal vector with mean $\bm$ and positive definite covariance matrix $\bM$
while given a matrix $\bX=[\bx_1, \ldots, \bx_M]\in\C^{N\times M}$, writing $\bX\sim\cC\cN_{N,M}(\bm,\bM,\bI)$ means
that $\bx_i\sim\cC\cN_N(\bm, \bM)$, $i=1,\ldots,M$, and the $\bx_i$s are statistically independent.

\section{Problem Formulation and Preliminary Definitions}
\label{Sec:Problem_Formulation}
Let us denote by $\bz_1,\ldots,\bz_K$ the $N$-dimensional vectors representing the returns
from $K$ range bins forming the region of interest illuminated by the radar system. The size $N$ of
such vectors represents the number of space, time,
or space-time channels. An important assumption of practical value
is that the statistical characterization of the clutter component is range-dependent \cite{Richards} and, hence, the corresponding
distribution parameters might change over the range (e.g., due to the presence of clutter boundaries).
Thus, the set $\Omega=\{1,\ldots,K\}$ of indices
can be partitioned into a given number, $L$ say, of subsets whose elements index data vectors sharing
the same statistical characterization
of the clutter; the $l$th subset is denoted by $\Omega_l = \{ {i_{l,1}},\ldots,{i_{l,K_l}} \}$,
where $K_l>N$, $l\in\ccL=\{1, \ldots, L\}$, denotes its cardinality.
The value of $L$ is assumed known and can be set exploiting the a priori information about the
terrain types composing the region of interest.
However, it is also likely that multiple point-like targets are present within the region of interest with
the consequence that data vectors indexed by $\Omega_l$, although homogeneous from the standpoint of clutter statistical
characterization, are heterogeneous due to target components
affecting $\bz_{i_{l,m}}$ for some $m\in\{1,\ldots,K_l\}$.
Summarizing, if data indexed by the generic $\Omega_l$ are free of target components, then
$\bZ_l=[\bz_{i_{l,1}}, \ldots, \bz_{i_{l,K_l}}] \sim \cC\cN_{N,K_l}(\bzero,\bM_l,\bI)$; when
there exists $\Omega_l^t=\{ {j_{l,1}}, \ldots, {j_{l,T_{l}}} \}\subseteq\Omega_l$
indexing vectors that contain target components with $T_l$ the number of targets within $\Omega_l$, then $\forall h=1,\ldots,T_l$,
\be
\bz_{j_{l,h}}\sim\cC\cN_N(\alpha_{j_{l,h}}\bv,\bM_l)
\label{eqn:dataTarget_S0}
\ee
for deterministic targets and
\be
\bz_{j_{l,h}}\sim\cC\cN_N(\bzero,\bM_l+\sigma^2_{j_{l,h}}\bv\bv^\dag)
\label{eqn:dataTarget_S1}
\ee
for fluctuating targets. In both cases, the $\bz_{j_{l,h}}$s are
statistically independent of $\bz_{i_{l,m}}\sim\cC\cN_N(\bzero,\bM_l)$,
$\forall i_{l,m}\neq j_{l,h}$, $m=1,\ldots,K_l$, $h=1,\ldots,T_l$.
In \eqref{eqn:dataTarget_S0} and \eqref{eqn:dataTarget_S1}, $\alpha_{j_{l,h}}\in\C$ and
$\sigma^2_{j_{l,h}}>0$ are the target response and the target power from the $j_{l,h}$th range bin
and along the nominal (space, time, or space-time) steering vector $\bv$ \cite{BOR-Morgan}, respectively.

From the above aspects, it turns out that the problem at hand has a twofold nature, namely it can be viewed as
\begin{itemize}
\item a classification problem of clutter returns;
\item a detection problem of multiple targets in heterogeneous clutter.
\end{itemize}
Thus, assuming that the clutter distribution parameters are unknown and, hence, must be estimated from data,
we can formulate such a problem in terms of a binary hypothesis test where the null hypothesis is given by
\be
H_0: \ \bz_i\sim\cC\cN_N(\bzero,\bM_l), \ i\in\Omega_l, \ l=1,\ldots, L,
\ee
the alternative hypothesis for nonfluctuating targets is
\be
H_{1,1}: \
\begin{cases}
\bz_i\sim\cC\cN_N(\bzero,\bM_l), \ i\in\Omega_l\setminus\Omega_l^t,
\\
\bz_i\sim\cC\cN_N(\alpha_i\bv,\bM_l), \ i\in\Omega_l^t,
\end{cases}
l=1,\ldots, L,
\label{eqn:swerling0}
\ee
and the alternative hypothesis for fluctuating targets can be expressed as
\be
H_{1,2}: \
\begin{cases}
\bz_i\sim\cC\cN_N(\bzero,\bM_l), \ i\in\Omega_l\setminus\Omega_l^t,
\\
\bz_i\sim\cC\cN_N(\bzero,\bM_l+\sigma^2_i \bv\bv^\dag), \ i\in\Omega_l^t,
\label{eqn:rankOneSteer}
\end{cases}
l=1,\ldots,L.
\ee
If we forget the structure of the target component, namely $\sigma^2_i \bv\bv^\dag$,
and assume that possible swarms of targets are present in the operating scenario,
the alternative hypothesis for random targets can be further
recast as
\be
H_{1,3}: \
\begin{cases}
\bz_i\sim\cC\cN_N(\bzero,\bM_l), \ i\in\Omega_l\setminus\Omega_l^t,
\\
\bz_i\sim\cC\cN_N(\bzero,\bM_l+\bR_l), \ i\in\Omega_l^t,
\end{cases}
l=1,\ldots, L,
\label{eqn:rankOneMatrix}
\ee
where $\bR_l\in\C^{N\times N}$ is a rank one matrix representative of the specific target swarm.
It is understood that each element of the swarm shares the same velocity and motion direction.
As for the power, actually, each target
should have its own power level and, hence, the index of the matrix representative of the target
should be $i\in\Omega_l^t$.  However, such a model leads to
intractable mathematics in the subsequent developments.
The same effect is also observed if we assume a common mean power for all targets, namely,
only one target covariance matrix $\bR$. For the above reasons, we follow an alternative route where
a swarm power class corresponds to a clutter covariance structure. This line of reasoning is motivated
by the fact that when the swarm is located in a clutter region close to the radar, its power would likely be
higher than that of an analogous fleet of targets in a clutter region far from the radar. As a consequence,
it is possible to associate a swarm power to each clutter covariance class.
Finally, estimating the unknown quantities under \eqref{eqn:swerling0},
\eqref{eqn:rankOneSteer}, and \eqref{eqn:rankOneMatrix} allows us to accomplish
the classification task.

Before concluding this section, we provide the expressions of the PDF that will be used in what follows
for the design of the detection architectures. Let us start by defining
$\cP_0^\prime=\{\Omega_l,\bM_l:l\in\ccL\}$,
$\cP^\prime_{1,1}=\{{\Omega_l,\alpha_i,\bM_l: \ i\in\Omega_l^t}, \ l\in\ccL\}$,
$\cP^\prime_{1,2}=\{\Omega_l, \sigma^2_i, \bM_l: \ i\in\Omega_l^t, \ l\in\ccL\}$,
and $\cP^\prime_{1,3}=\{\Omega_l,\Omega_l^t,\bM_l, \bR_l: \ l\in\ccL\}$ the sets of unknown parameters
under $H_0$, $H_{1,1}$, $H_{1,2}$, and $H_{1,3}$, respectively.
The joint PDF of $\bZ=[\bz_1,\ldots,\bz_K]$ under $H_0$ is
\be
f_0(\bZ; \cP_0^\prime )=\prod_{l=1}^L\prod_{i\in\Omega_l}
\frac{\exp\{ -\tr[\bM_l^{-1}\bz_i\bz_i^\dag] \}}
{\pi^N\det(\bM_l)};
\ee
under $H_{1,1}$ it can be written as
\begin{multline}
f_1(\bZ; \cP_{1,1}^\prime) = \prod_{l=1}^L \Bigg[ \prod_{i\in\Omega_l\setminus\Omega_l^t}
\frac{\exp\{ -\tr[\bM_l^{-1}\bz_i\bz_i^\dag] \}}
{\pi^N\det(\bM_l)}
\\
\times\prod_{i\in \Omega_l^t}
\frac{\exp\{ -\tr[\bM_l^{-1}(\bz_i-\alpha_i\bv)(\bz_i-\alpha_i\bv)^\dag] \}}
{\pi^N\det(\bM_l)}\Bigg];
\end{multline}
under $H_{1,2}$, it is given by
\[
f_1(\bZ; \cP_{1,2}^\prime) = \prod_{l=1}^L \Bigg[ \prod_{i\in\Omega_l\setminus\Omega_l^t}
\frac{\exp\{ -\tr[\bM_l^{-1}\bz_i\bz_i^\dag] \}}
{\pi^N\det(\bM_l)}
\]
\be
\times\prod_{i\in \Omega_l^t}
\frac{\exp\{ -\tr[(\bM_l + \sigma^2_i\bv\bv^\dag)^{-1}\bz_i\bz_i^\dag] \}}
{\pi^N\det(\bM_l+ \sigma^2_i\bv\bv^\dag)}\Bigg];
\ee
finally, that related to $H_{1,3}$ is
\begin{multline}
f_1(\bZ; \cP_{1,3}^\prime) = \prod_{l=1}^L \Bigg[ \prod_{i\in\Omega_l\setminus\Omega_l^t}
\frac{\exp\{ -\tr[\bM_l^{-1}\bz_i\bz_i^\dag] \}}
{\pi^N\det(\bM_l)}
\\
\times\prod_{i\in \Omega_l^t}
\frac{\exp\{ -\tr[(\bM_l + \bR_l)^{-1}\bz_i\bz_i^\dag] \}}
{\pi^N\det(\bM_l + \bR_l)}\Bigg].
\end{multline}

\section{Architecture Designs and Estimation Procedures}
\label{Sec:Architecture_Designs}
The architectures devised in this section are grounded on the LRT where the unknown parameters are
replaced by suitable estimates. Thus, denoting by $\widehat{\bR}_l$,
$\widehat{\Omega}_l$, $\widehat{\Omega}_l^t$,
$\widehat{\bM}_l$, $l=1,\ldots,L$, along with $\widehat{\sigma}^2_m$
and $\widehat{\alpha}_m$, $m\in\widehat{\Omega}_l^t$,
the estimates of the respective unknown parameters, the LRT for
testing $H_0$ against $H_{1,i}$ has the following form
\be
\frac{f_1(\bZ; \widehat{\cP^\prime}_{1,i})}
{f_0(\bZ; \widehat{\cP^\prime}_{0})}\test \eta,
\label{eqn:LRT_1}
\ee
where $\widehat{\cP^\prime}_{1,i}$, $i=1,2,3$, and $\widehat{\cP^\prime}_{0}$ are the sets
of estimates under $H_{1,i}$, $i=1,2,3$, and $H_0$,
respectively; $\eta$ is the threshold\footnote{Hereafter, $\eta$ denotes the generic
detection threshold.} to be set according to the required probability of false alarm ($P_{fa}$).
Now, exploiting the MLA to come up with $\widehat{\cP^\prime}_{1,i}$ and $\widehat{\cP^\prime}_{0}$ does not
seem viable from the standpoints of mathematical tractability and computational load. In fact, this approach
requires to account for all the range bin configurations in terms of presence
of targets and/or specific clutter covariance components. For this reason, we extend the methodology proposed in
\cite{clutterClust} to the case contemplating the presence of targets and introduce
hidden discrete random variables that are representative of the
covariance class associated with a given range bin as well as of the presence of a target with a given
Angle of Arrival (AoA) and/or normalized Doppler frequency in that range bin.

Specifically, under the alternative hypotheses, let us assume that $K$ independent and identically
distributed discrete random variables, $c_k$s say, are associated with the range bins under test
and that they are not observable. Such random variables have unknown PMF
$P(c_k=l) = p_l$, $k=1,\ldots,K$ and
$l\in\{1,\ldots,L_c\}$.  As better explained below, $L_c$ accounts for the number of clutter covariance classes and
the presence of a possible target.
Thus, we can use the index $l$ to code different operating
situations. Specifically, under the alternative hypotheses, when $c_k=l$ and
$Ls + 1\leq l \leq Ls+L$, $s=0,1$, then
\be
\bz_k\sim
\begin{cases}
\cC\cN_N(\alpha_{s,k} \bv,\bM_{l-Ls}),  \mbox{ under $H_{1,1}$},
\\
\cC\cN_N(\bzero,\bM_{l-Ls}+\sigma^2_{s,k}\bv\bv^\dag), \mbox{ under $H_{1,2}$},
\\
\cC\cN_N(\bzero,\bM_{l-Ls}+s\bR_{l-Ls}), \mbox{ under $H_{1,3}$},
\end{cases}
\ee
where we set $\alpha_{0,k}=\sigma^2_{0,k}=0$.
On the other hand, when $H_0$ is in force, we assume that
$c_k=l$, $l\in\ccL$, implies that $\bz_k\sim\cC\cN_N(\bzero,\bM_l)$.
As a consequence, $L_c=L$ under $H_0$ and $L_c=2L$ under $H_{1,i}$,
$i=1,2,3$.\footnote{Recall that $\sum_{s=0}^{1}\sum_{l=1}^L p_{Ls+l}=1$
under $H_{1,i}$, $i=1,2,3$.}
The corresponding sets of unknown parameters associated with the distribution of $\bz_k$
are $\cP_{0}=\cP_{0}^\prime \cup \cA$ with
$\cA=\{ p_1,\ldots,p_{L_c} \}$ under $H_0$,
$\cP_{1,1,k}=\cA \cup \{\alpha_{1,k}, \bM_l: l\in\ccL\}$
under $H_{1,1}$, $\cP_{1,2,k}=\cA \cup \{\sigma^2_{1,k},\bM_l: l\in\ccL\}$
under $H_{1,2}$, and $\cP_{1,3,k}=\cA \cup \{\bM_l, \bR_l: \ l\in\ccL \}$ under $H_{1,3}$.

With the above remarks in mind, we rewrite the PDF of $\bz_k$ under $H_{1,i}$, $i=1,2,3$, as
\be
g_1(\bz_k;\cP_{1,i,k}) = \sum_{s=0}^{1}\sum_{l=1}^L p_{Ls+l} f(\bz_k|c_k=Ls+l;\Theta_{s,i,k,l}),
\label{eqn:pdf_LVM}
\ee
where
\be
\Theta_{s,i,k,l}=
\begin{cases}
\{\alpha_{1,k},\bM_l\}, & \mbox{for } s=1, \ i=1,
\\
\{\sigma^2_{1,k},\bM_l\},  & \mbox{for } s=1, \ i=2,
\\
\{\bR_l,\bM_l\},  & \mbox{for } s=1, \ i=3,
\\
\{ \bM_l \}, & \mbox{for } s=0, \ i=1,2,3.
\end{cases}
\ee
As for the PDF of $\bz_k$ under $H_0$, it can be written
as in \cite{clutterClust} by setting $s=0$, $\forall l=1,\ldots,L$, and, hence,
neglecting the outer summation in the previous PDFs, namely,
\be
g_0(\bz_k;\cP_0) = \sum_{l=1}^L p_l f(\bz_k|c_k=l;\bM_l).
\label{eqn:pdf_LVM_H0}
\ee
In \eqref{eqn:pdf_LVM}, $f(\bz_k|c_k=Ls+l;\Theta_{s,1,k,l})$
denotes the PDF of a complex Gaussian vector with covariance matrix $\bM_l$ and
mean $\alpha_{1,k}\bv$ if $s=1$ whereas the mean is zero for $s=0$;
$f(\bz_k|c_k=Ls+l;\Theta_{s,2,k,l})$ is the PDF of a complex Gaussian vector with zero mean and covariance matrix
$\bM_l+\sigma^2_{1,k}\bv\bv^\dag$ if $s=1$, when $s=0$ the covariance matrix is
$\bM_l$; $f(\bz_k|c_k=Ls+l;\Theta_{s,3,k,l})$ is the PDF of a complex Gaussian vector with zero mean and covariance matrix
$\bM_l+s\bR_l$; finally, in \eqref{eqn:pdf_LVM_H0},
$f(\bz_k|c_k=l;\bM_l)$ is the multivariate complex Gaussian PDF with zero mean
and covariance matrix $\bM_l$.
Exploiting the above characterizations, given $i=1,2,3$, we can also build up the following alternative test
\be
\prod_{k=1}^K \frac{g_1(\bz_k;\widehat{\cP}_{1,i,k})}
{g_0(\bz_k;\widehat{\cP}_0)}\test \eta.
\label{eqn:LRT_LVM}
\ee
Notice that also in this case, obtaining the maximum likelihood estimates
of the unknown parameters is a difficult task. However, the presence of the hidden random variables
allows us to solve the estimation problems under all the considered hypotheses by devising iterative procedures based
upon the EM algorithm \cite{Dempster77}. More importantly,
such estimates can be exploited to obtain data partitions that can also be used in \eqref{eqn:LRT_1}.

The estimation procedures under $H_0$ have been already developed in \cite{clutterClust} and, hence,
we focus on those under $H_{1,i}$, $i=1,2,3$.

\subsection{Estimation procedures for deterministic targets ($H_{1,1}$)}
In order to apply the EM algorithm under $H_{1,1}$, let us exploit \eqref{eqn:pdf_LVM}
with $i=1$ and write the joint
log-likelihood of $\bZ$ for deterministic targets as follows
\begin{align*}
&\cL(\bZ; \cP_{1,1}) = \sum_{k=1}^K \log g_1(\bz_k;\cP_{1,1,k}) \nonumber
\\
&=\sum_{k=1}^K \log \left[
\sum_{s=0}^{1}\sum_{l=1}^L p_{Ls+l} f(\bz_k|c_k=Ls+l;\Theta_{s,1,k,l})
\right],
\end{align*}
where $\cP_{1,1}=\cup_{k=1}^K\cP_{1,1,k}$.\footnote{In what follows, we denote
by $\cL(\bZ; \cP_{1,i})$ with $\cP_{1,i}=\cup_{k=1}^K\cP_{1,i,k}$, $i=1,2,3$, the joint
log-likelihood function under $H_{1,i}$.}
The first step of the EM is the E-step that leads to the following update rule
\begin{align}
&q_k^{(h-1)}(Ls+l)=p\left( c_k=Ls+l |  \bz_k; \widehat{\cP}_{1,1,k}^{(h-1)}\right) \nonumber
\\
&= \frac{\ds f\left( \bz_k|c_k=Ls+l; \widehat{\Theta}_{s,1,k,l}^{(h-1)} \right) \widehat{p}_{Ls+l}^{(h-1)} }
{\ds \sum_{j=0}^{1}\sum_{i=1}^L
f\left( \bz_k|c_k=Lj+i;\widehat{\Theta}_{j,1,k,i}^{(h-1)} \right) \widehat{p}_{Lj+i}^{(h-1)} },
\label{eqn:E-step_}
\end{align}
$s=0,1$, where $\widehat{\Theta}_{s,1,k,l}^{(h-1)}$ and $\widehat{p}_{Ls+l}^{(h-1)}$
are the estimates of the unknown parameters
at the previous step.\footnote{In what follows, the estimate of a parameter vector $\bp$ at the $h$th step
is denoted by $\widehat{\bp}^{(h)}$.}
In what follows, in order to mitigate the inclination
of the likelihood function to select the hypotheses that include target components,
we borrow the likelihood approximations that give rise to the Model Order Selection rules \cite{Stoica1}
and write \eqref{eqn:E-step_} as follows
\be
\mbox{\eqref{eqn:E-step_}}\!\approx\!
\frac{\ds f\!\left(\! \bz_k|c_k\!=\!Ls\!+\!l;\widehat{\Theta}_{s,1,k,l}^{(h-1)} \!\right)\!
e^{-u(s)} \widehat{p}_{Ls+l}^{(h-1)} }
{\ds \sum_{j=0}^{1}\sum_{i=1}^L
f\!\left( \!\bz_k|c_k\!=\!Lj\!+\!i;\widehat{\Theta}_{j,1,k,i}^{(h-1)} \!\right)\!
e^{-u(j)} \widehat{p}_{Lj+i}^{(h-1)} },
\label{eqn:E-step}
\ee
where, given the PDF of $\bz_k$, $u(j)$, $j=0,1$, is a penalty function depending on the number of
the unknown parameters (a point better explained in Section \ref{Sec:performance}).

The second step is the M-step that requires to solve the following problem
\begin{multline}
\widehat{\cP}_{1,1}^{(h)}=\argmax_{\cP_{1,1}}\Bigg\{
\sum_{k=1}^K \sum_{s=0}^{1} \sum_{l=1}^L q_k^{(h-1)}(Ls+l)
\\
\times \log \frac{ f(\bz_k|c_k=Ls+l;{\Theta}_{s,1,k,l}) p_{Ls+l}}
{q_k^{(h-1)}(Ls+l)} \Bigg\}
\end{multline}
that is tantamount to
\begin{align}
\widehat{\cP}_{1,1}^{(h)}&=\argmax_{\cP_{1,1}}
\left\{
\sum_{k=1}^K \sum_{s=0}^{1} \sum_{l=1}^L  q_k^{(h-1)} (Ls+l) \right.\nonumber
\\
&\times \log f(\bz_k|c_k=Ls+l;{\Theta}_{s,1,k,l})  \nonumber
\\
&
\left.+\sum_{k=1}^K \sum_{s=0}^{1} \sum_{l=1}^L q_k^{(h-1)}(Ls+l) \log p_{Ls+l}
\right\}.
\end{align}
Now, we proceed with the maximization step and observe that
the optimization problem with respect to $p_{Ls+l}$, $s=0,1$, $l=1,\ldots,L$, is independent of
that over $\bM_l$ and $\alpha_{1,k}$.
Thus, we can proceed by separately addressing these two problems.
Starting from the optimization over $\cA$, observe that it can be solved by using
the method of Lagrange multipliers to take into account the constraint
$\sum_{s=0}^{1}\sum_{l=1}^L p_{Ls+l}=1$. The final result is
\be
\widehat{p}_{Ls+l}^{(h)}= \frac{1}{K} \sum_{k=1}^K q_k^{(h-1)}(Ls+l), \ s=0,1, \ l=1,\ldots,L.
\ee
Thus, we can focus on the maximization problem with respect
to the $\alpha_{1,k}$s and $\bM_l$s, i.e.,
\begin{multline}
\dmax_{\alpha_{1,k}\atop k=1,\ldots,K}\dmax_{{\bf M}_l \atop l=1,\ldots,L}
\sum_{k=1}^K \sum_{s=0}^{1} \sum_{l=1}^L  q_k^{(h-1)} (Ls+l)
\\
\times \log f(\bz_k|c_k=Ls+l;{\Theta}_{s,1,k,l}).
\end{multline}
Replacing the PDFs with their expressions and ignoring the terms independent of the parameters of interest, the above problem
is tantamount to
\begin{align}
\dmin_{\alpha_{1,k} \atop k=1,\ldots,K}
&\dmin_{{\bf M}_l \atop l=1,\ldots,L}
\sum_{k=1}^K \sum_{s=0}^{1} \sum_{l=1}^L q_k^{(h-1)} (Ls+l)\Bigg\{
\log\det(\bM_l)
\nonumber
\\
&+\tr\left[\bM_l^{-1}(\bz_k-\alpha_{s,k}\bv)(\bz_k-\alpha_{s,k}\bv)^\dag\right]
\Bigg\}.
\label{eqn:originalProblem}
\end{align}
A suboptimum solution to this optimization can be found by resorting to a cyclic procedure that repeats
the following steps until a stopping criterion is not satisfied
\begin{itemize}
\item assume that the $\alpha_{1,k}$s are known and estimate $\bM_l$;
\item set the $\bM_l$s to the values obtained at the previous step and estimate the $\alpha_{1,k}$s.
\end{itemize}
Thus, denote by $\widehat{\alpha}_{1,k}^{(h-1),(m-1)}$, $k=1,\ldots,K$, the estimates of the $\alpha_{1,k}$s
at the $(h-1)$th EM step and $(m-1)$th step of this inner
procedure (we set $\widehat{\alpha}_{0,k}^{(h-1),(m-1)}=0$),
then we solve the following problem
\begin{align}
&\dmin_{{\bf M}_l}\left\{
-\log\det(\bM_l^{-1}) \sum_{k=1}^K\sum_{s=0}^{1} q_k^{(h-1)}(Ls+l)\right. \nonumber
\\
&+\tr\Bigg[\bM_l^{-1} \sum_{k=1}^K\sum_{s=0}^{1}q_k^{(h-1)}(Ls+l) \nonumber
\\
&\times \left.(\bz_k-\widehat{\alpha}_{s,k}^{(h-1),(m-1)}\bv)
(\bz_k-\widehat{\alpha}_{s,k}^{(h-1),(m-1)}\bv)^\dag\Bigg]\right\}.
\end{align}
The minimizer can be obtained by resorting to the following inequality \cite{lutkepohl1997handbook}
$\log \det(\bA) \leq \tr[\bA] - N$,
where $\bA$ is any $N$-dimensional matrix with nonnegative eigenvalues, and, hence, we come up with
\begin{multline}
\widehat{\bM}_l^{(h-1),(m)}=\frac{1}
{\ds\sum_{s=0}^{1}\sum_{k=1}^K q_k^{(h-1)}(Ls+l)}
\sum_{s=0}^{1}\sum_{k=1}^K q_k^{(h-1)}(Ls+l)
\\
\times
(\bz_k-\widehat{\alpha}_{s,k}^{(h-1),(m-1)}\bv)(\bz_k-\widehat{\alpha}_{s,k}^{(h-1),(m-1)}\bv)^\dag.
\end{multline}
Now, assuming that in \eqref{eqn:originalProblem} $\bM_l=\widehat{\bM}_l^{(h-1),(m)}$, $l=1,\ldots,L$,
we obtain
\begin{multline}
\dmin_{\alpha_{1,k} \atop k=1,\ldots,K}
\sum_{k=1}^K \sum_{l=1}^L q_k^{(h-1)}(L+l)
(\bz_k-\alpha_{1,k}\bv)^\dag
\\
\times \left(\widehat{\bM}_l^{(h-1),(m)}\right)^{-1} (\bz_k-\alpha_{1,k}\bv),
\end{multline}
and setting to zero the first derivative of the objective function
with respect to $\alpha_{1,k}$ leads to
\be
\widehat{\alpha}_{1,k}^{(h-1),(m)}=
\frac{\ds\sum_{l=1}^L q_k^{(h-1)}(L+l) \bv^\dag \left(\widehat{\bM}_l^{(h-1),(m)}\right)^{-1}\bz_k}
{\ds\sum_{l=1}^L q_k^{(h-1)}(L+l) \bv^\dag \left(\widehat{\bM}_l^{(h-1),(m)}\right)^{-1}\bv}.
\ee
This inner procedure continues until a convergence criterion is not satisfied. The final estimates
of $\alpha_{1,k}$ and $\bM_l$ are denoted by $\widehat{\alpha}_{1,k}^{(h)}$ and $\widehat{\bM}_l^{(h)}$, respectively,
and are used in the next cycle of the EM-based procedure.
Before moving on, notice that the composition of the EM and this alternating procedure yields a nondecreasing
sequence of log-likelihood values.

\subsection{Estimation procedures for fluctuating targets ($H_{1,2}$ and $H_{1,3}$)}
In this subsection, we consider the alternative hypotheses defined by \eqref{eqn:rankOneSteer}
and \eqref{eqn:rankOneMatrix} as well as the related PDFs in the steps
of the EM algorithm. Starting from the expectation step and following the same line of reasoning as
for $H_{1,1}$, the final result maintains the form of \eqref{eqn:E-step_} and
its approximation \eqref{eqn:E-step} but for the
PDF of $\bz_k$. More precisely, replacing $\widehat{\Theta}_{s,1,k,l}^{(h-1)}$ in
\eqref{eqn:E-step} with $\widehat{\Theta}_{s,2,k,l}^{(h-1)}$ for model \eqref{eqn:rankOneSteer} and
with $\widehat{\Theta}_{s,3,k,l}^{(h-1)}$ for model \eqref{eqn:rankOneMatrix}, we come up with the respective
E-steps. Let us recall here that $\widehat{\Theta}_{s,2,k,l}^{(h-1)}$ and
$\widehat{\Theta}_{s,3,k,l}^{(h-1)}$ contain the
estimates of the unknown parameters under $H_{1,2}$ and $H_{1,3}$, respectively,
at the previous step.

As for the maximization steps, we separately proceed as described in the next subsections.

\subsubsection{Fluctuating targets model \eqref{eqn:rankOneSteer}}
The problem to be solved has the following expression
\begin{align}
&\dmin_{\sigma^2_{1,k} \atop k=1,\ldots,K}
\dmin_{{\bf M}_l \atop l=1,\ldots,L}
\sum_{k=1}^K \sum_{s=0}^{1} \sum_{l=1}^L q_k^{(h-1)} (Ls+l) \nonumber
\\
&\times \Bigg\{
\log\det(\bM_l+\sigma^2_{s,k}\bV)
+\tr\left[(\bM_l+\sigma^2_{s,k}\bV)^{-1}\bS_k\right]
\Bigg\},
\label{eqn:originalProblem_s1}
\end{align}
where $\bV=\bv\bv^\dag$ and $\bS_k=\bz_k \bz_k^\dag$.
The joint maximization with respect to the $\sigma^2_{s,k}$s and the $\bM_l$s is difficult from the standpoint
of mathematics (at least to the best of authors' knowledge). For this reason, we exploit a heuristic approach
that leads to suitable solutions. Specifically, let us denote the objective function in \eqref{eqn:originalProblem_s1}
by $x(\cP_{1,2})$ with $\cP_{1,2}=\cup_{k=1}^K\cP_{1,2,k}$ and observe that it can be recast as
$x(\cP_{1,2})=x_1(\bM_1,\ldots,\bM_L)+x_2(\cP_{1,2})$, where
$x_1(\bM_1,\ldots,\bM_L)=\sum_{k=1}^K \sum_{l=1}^L q_k^{(h-1)} (l) \{\log\det(\bM_l)
+\tr\left[\bM_l^{-1}\bS_k\right]\}$
and
$x_2(\cP_{1,2})\!=\!\!\!\sum_{k=1}^K \sum_{l=1}^L \! q_k^{(h-1)} (L+l) \{
\log\det(\bM_l \!+\! \sigma^2_{1,k}\bV)+\tr[(\bM_l+\sigma^2_{1,k}\bV)^{-1}\bS_k]\}$.
Since $x_1(\cdot)$ depends on the covariance matrices only, we first
estimate $\bM_l$, $l=1,\ldots,L$, by exploiting $x_1(\cdot)$, then replace such estimates in
$x_2(\cdot)$, and minimize it with respect to $\sigma^2_{1,k}$, $k=1,\ldots,K$.
Thus, the estimates of $\bM_l$, $l=1,\ldots,L$, based upon $x_1(\cdot)$ are given by
\be
\widehat{\bM}_l^{(h)}=
\left\{\ds \sum_{k=1}^K q_k^{(h-1)} (l)\right\}^{-1} \sum_{k=1}^K q_k^{(h-1)} (l) \bS_k.
\ee
Now, we consider $x_2(\cdot)$ and set $\bM_l=\widehat{\bM}_l^{(h)}$, $l=1,\ldots,L$. Then,
the problem at hand becomes
\begin{multline}
\dmin_{\sigma^2_{1,k} \atop k=1,\ldots,K}
\sum_{k=1}^K \sum_{l=1}^L \! q_k^{(h-1)} (L+l)
\Bigg\{ \!\!
\log\det\left(\widehat{\bM}_l^{(h)} \!+\! \sigma^2_{1,k}\bV\right)
\\
+\tr\left[\left(\widehat{\bM}_l^{(h)}+\sigma^2_{1,k}\bV\right)^{-1}\bS_k\right]
\Bigg\}.
\label{eqn:min_sigma}
\end{multline}
In order to solve \eqref{eqn:min_sigma}, we consider a given $k$ and the related objective function
\begin{align}
&h_{k}(\sigma^2_{1,k})=\sum_{l=1}^L \! q_k^{(h-1)} (L+l)
\nonumber
\\
&\times\Bigg\{ \!\!
\log\det\left(\widehat{\bM}_l^{(h)} \!+\! \sigma^2_{1,k}\bV\right)
\nonumber
\\
&+\tr\left[\left(\widehat{\bM}_l^{(h)}+\sigma^2_{1,k}\bV\right)^{-1}\bS_k\right]
\Bigg\}
\nonumber
\\
&=\sum_{l=1}^L \! q_k^{(h-1)} (L+l) \log\left[1+\sigma^2_{1,k}\bv^\dag
(\widehat{\bM}_l^{(h)})^{-1}\bv\right]
\nonumber
\\
&+\sum_{l=1}^L \! q_k^{(h-1)} (L+l)\log\det\left(\widehat{\bM}_l^{(h)}\right)
\nonumber
\\
&-\sum_{l=1}^L \! q_k^{(h-1)} (L+l)\sigma^2_{1,k}\frac{|\bz_k^\dag (\widehat{\bM}_l^{(h)})^{-1} \bv|^2}
{1+\sigma^2_{1,k}\bv^\dag (\widehat{\bM}_l^{(h)})^{-1}\bv}
\nonumber
\\
&+\sum_{l=1}^L \! q_k^{(h-1)} (L+l)\bz_k^\dag (\widehat{\bM}_l^{(h)})^{-1} \bz_k.
\label{eqn:min_sigma_2}
\end{align}
Thus, setting to zero the first derivative with respect to $\sigma^2_{1,k}$, we obtain
\begin{align}
&\sum_{l=1}^L \frac{q_k^{(h-1)} (L+l) \bv^\dag (\widehat{\bM}_l^{(h)})^{-1}\bv}
{1+\sigma^2_{1,k}\bv^\dag (\widehat{\bM}_l^{(h)})^{-1}\bv}
\nonumber
\\
&-\sum_{l=1}^L \frac{q_k^{(h-1)} (L+l) |\bz_k^\dag (\widehat{\bM}_l^{(h)})^{-1} \bv|^2}
{1+\sigma^2_{1,k}\bv^\dag (\widehat{\bM}_l^{(h)})^{-1}\bv}
\nonumber
\\
&+\sum_{l=1}^L  \frac{\sigma^2_{1,k} q_k^{(h-1)} (L+l)|\bz_k^\dag (\widehat{\bM}_l^{(h)})^{-1} \bv|^2}
{\left(1+\sigma^2_{1,k}\bv^\dag (\widehat{\bM}_l^{(h)})^{-1}\bv\right)^2
\left(\bv^\dag (\widehat{\bM}_l^{(h)})^{-1}\bv\right)^{-1}}=0
\label{eqn:derivative}
\end{align}
namely
\be
\sum_{l=1}^L \frac{q_k^{(h-1)} (L+l) [\sigma^2_{1,k}(a_{l}^{(h)})^2+a_{l}^{(h)}-b_{k,l}^{(h)}]  }
{\left(1+\sigma^2_{1,k}a_{l}^{(h)}\right)^2}=0,
\ee
where $a_{l}^{(h)}=\bv^\dag (\widehat{\bM}_l^{(h)})^{-1}\bv$ and
$b_{k,l}^{(h)}=|\bz_k^\dag (\widehat{\bM}_l^{(h)})^{-1} \bv|^2$.
The solutions of the last equation can be found by means of numerical routines and, then, we choose the positive
solution (if any) that minimizes \eqref{eqn:min_sigma_2}.

\subsubsection{Fluctuating targets model \eqref{eqn:rankOneMatrix}}
Let us assume that $H_{1,3}$ (see equation \eqref{eqn:rankOneMatrix}) is true and
recall that $\bR_k$ is positive semidefinite of rank $1$.
The maximization step consists in solving
\begin{multline}
\dmin_{ {{\bf M}_l, {\bf R}_l} \atop l=1,\ldots,L}
\sum_{k=1}^K \sum_{s=0}^{1} \sum_{l=1}^L q_k^{(h-1)} (Ls+l)
 \Bigg\{
\log\det(\bM_l+s\bR_l)
\\
+\tr\left[(\bM_l+s\bR_l)^{-1}\bz_k \bz_k^\dag\right]
\Bigg\}.
\label{eqn:originalProblem_m1}
\end{multline}
Notice that the objective function can be further rewritten as
\begin{multline}
\sum_{k=1}^K \sum_{l=1}^L q_k^{(h-1)} (l) \Bigg\{
\log\det(\bM_l)
+\tr\left[\bM_l^{-1}\bz_k \bz_k^\dag\right]
\Bigg\}
\\
+\sum_{k=1}^K \sum_{l=1}^L q_k^{(h-1)} (L+l)\Bigg\{
\log\det(\bM_l+\bR_l)
\\
+\tr\left[(\bM_l+\bR_l)^{-1}\bz_k \bz_k^\dag\right]
\Bigg\}.
\label{eqn:objectiveFunction}
\end{multline}
Thus, in order to minimize the above function with respect to $\bM_l$ and $\bR_l$, let us consider the generic
index $l$ and the corresponding term
\begin{multline}
q_l\log\det(\bM_l)  + \tr\left[\bM_l^{-1}\bS_l \right]
\\
+q_{l+L}\log\det(\bM_l+\bR_l)
+\tr\left[(\bM_l+\bR_l)^{-1}\bS_{l+L}\right],
\label{eqn:objectiveFunction_l}
\end{multline}
where $q_l=\sum_{k=1}^K q_k^{(h-1)} (l)$, $\bS_l=\sum_{k=1}^K q_k^{(h-1)} (l)\bz_k \bz_k^\dag$,
$q_{l+L}=\sum_{k=1}^K q_k^{(h-1)} (L+l)$, and $\bS_{l+L}=\sum_{k=1}^K q_k^{(h-1)} (l+L)\bz_k \bz_k^\dag$.
Then, define $\bA_l=\bM_l^{1/2}\bU_l$ with $\bU_l\in\C^{N\times N}$ the unitary matrix whose columns are the
eigenvectors of $\bM_l^{-1/2} (\bM_l+\bR_l) \bM_l^{-1/2}$ and $\bLambda_l=\diag(\lambda_{l,1},1,\ldots,1)$,
$\lambda_{l,1}>1$,
the diagonal matrix containing the eigenvalues of $\bM_l^{-1/2} (\bM_l+\bR_l) \bM_l^{-1/2}$. It clearly
follows that $\bM_l=\bA_l\bA_l^\dag$, $(\bM_l+\bR_l)=\bA_l\bLambda_l\bA_l^\dag$, and
\eqref{eqn:objectiveFunction_l} can be written as
\begin{multline}
2(q_l+q_{l+L})\log|\det(\bA_l)|+\tr[\bA_l^{-1}\bS_l\bA^{-\dag}]
\\
+q_{l+L}\log\det(\bLambda_l)+\tr[\bLambda_l^{-1}\bA_l^{-1}\bS_{l+L}\bA_l^{-\dag}].
\label{eqn:objectiveFunction_Al}
\end{multline}
Now, let us exploit the eigendecomposition of $\bS_l^{-1/2} \bS_{l+L} \bS_l^{-1/2}=\bV_l\bGamma_l\bV_l^\dag$ where
$\bGamma_l=\diag(\gamma_{l,1},\ldots,\gamma_{l,N})$, $\gamma_{l,1}\geq\ldots\geq\gamma_{l,N}\geq 0$, is the
diagonal matrix of the eigenvalues and $\bV_l\in\C^{N\times N}$ is the unitary matrix of the corresponding eigenvectors.
Denoting by $\bB_l=\bS_l^{1/2}\bV_l$, we obtain that $\bS_l=\bB_l\bB_l^\dag$, $\bS_{l+L}=\bB_l\bGamma_l\bB_l^\dag$, and
\eqref{eqn:objectiveFunction_Al} can be recast as
\begin{multline}
c_l\left[\log|\det(\bB_l)|-(1/2)\log\det(\bLambda_l)-\log|\det(\bX_l)|\right]
\\
+\tr[\bX_l^\dag\bLambda_l\bX_l]+ q_{l+L}\log\det(\bLambda_l) + \tr[\bX_l\bGamma_l\bX_l^\dag],
\end{multline}
where $c_l=2(q_l+q_{l+L})$ and $\bX_l=\bLambda_l^{-1/2}\bA_l^{-1}\bB_l$. Exploiting the singular value decomposition
of $\bX_l$ given by $\bX_l=\bT_l\bD_l\bQ_l$, $\bD_l=\diag(d_{l,1},\ldots,d_{l,N})$ and
$d_{l,1}\leq\ldots\leq d_{l,N}$, the last equation becomes
\begin{multline}
c_l\left[\log|\det(\bB_l)|-(1/2)\log\det(\bLambda_l)-\log\det(\bD_l)\right]
\\
+\tr[\bD^2_l\bT_l^\dag\bLambda_l\bT_l]+ q_{l+L}\log\det(\bLambda_l) + \tr[\bD^2_l\bQ_l\bGamma_l\bQ_l^\dag].
\end{multline}
By {\em Theorem 1} of \cite{mirsky1959trace}, the minimization with respect to $\bT_l$ and $\bQ_l$ leads to
\begin{align}
& \ c_l\left[\log|\det(\bB_l)|-\frac{1}{2}\log\det(\bLambda_l)-\log\det(\bD_l)\right]
\nonumber
\\
& +\tr[\bD^2_l\bLambda_l]+ q_{l+L}\log\det(\bLambda_l) + \tr[\bD^2_l\bGamma_l].
\nonumber
\\
&=c_l\log|\det(\bB_l)|+a_l\log\lambda_{l,1} +d^2_{l,1}(\lambda_{l,1}+\gamma_{l,1})
\nonumber
\\
& +\sum_{i=2}^N d^2_{l,i}(1+\gamma_{l,i})
-\frac{c_l}{2}\sum_{i=1}^N\log d^2_{l,i},
\label{eqn:objectiveFunction_DLambda}
\end{align}
where $a_l=-c_l/2+q_{l+L}$. Setting to zero the first derivative of the above function with respect to
$d_{l,i}^2$, we obtain
\be
\hat{d}_{l,i}^2=
\begin{cases}
\ds
\frac{c_l/2}{\lambda_{l,1}+\gamma_{l,1}}, & \mbox{if } i=1,
\\
\vspace{-3mm}
\\
\ds
\frac{c_l/2}{1+\gamma_{l,i}}, & \mbox{if } i>1.
\end{cases}
\ee
Finally, replacing $d_{l,i}^2$ with $\hat{d}_{l,i}^2$ in \eqref{eqn:objectiveFunction_DLambda} yields
\begin{multline}
c_l\log|\det(\bB_l)|+a_l\log\lambda_{l,1}+N\frac{c_l}{2}
\\
-\frac{c_l}{2}\log\frac{c_l/2}{\lambda_{l,1}+\gamma_{l,1}}-\frac{c_l}{2}\sum_{i=2}^N\log \frac{c_l/2}{1+\gamma_{l,i}}
\end{multline}
and setting to zero
the first derivative of the last equation with respect to $\lambda_{l,1}$, the result is
\be
\hat{\lambda}_{l,1}=\dmax\left\{ \frac{-2 a_l\gamma_{l,1}}{2a_l+c_l}, 1 \right\}
=\dmax\left\{ \frac{ q_l\gamma_{l,1}}{q_{l+L}}, 1 \right\}.
\ee

\subsection{Classification rule}
Once the unknown quantities have been estimated, data
classification can be accomplished by exploiting suitable rules.
Under $H_0$, we use the same rule as in \cite{clutterClust}, whereas under the alternative hypotheses
$\forall k=1,\ldots,K$, we proceed as follows
\be
\bz_k\sim
\begin{cases}
\cC\cN_N(\bzero,\widehat{\bM}^{(h_{\max})}_{\hat{l}_k}), & 1\leq\hat{l}_k\leq L,
\\
\cC\cN_N(\widehat{\alpha}_k^{(h_{\max})}{\bv},\widehat{\bM}^{(h_{\max})}_{\hat{l}_k-L}), & L+1\leq\hat{l}_k\leq 2L,
\end{cases}
\ee
under $H_{1,1}$,
\be
\bz_k\sim
\begin{cases}
\cC\cN_N\left(\bzero,\widehat{\bM}^{(h_{\max})}_{\hat{l}_k}\right), & \!\!\! 1\leq\hat{l}_k\leq L,
\\
\cC\cN_N(\bzero,\widehat{\bM}^{(h_{\max})}_{\hat{l}_k-L}+(\widehat{\sigma}^2_{1,k})^{(h_{\max})}\bv\bv^\dag), &
\!\!\! L+1\leq\hat{l}_k\leq 2L,
\end{cases}
\ee
under $H_{1,2}$, and
\be
\bz_k\sim
\begin{cases}
\cC\cN_N\left(\bzero,\widehat{\bM}^{(h_{\max})}_{\hat{l}_k}\right), & 1\leq\hat{l}_k\leq L,
\\
\cC\cN_N(\bzero,\widehat{\bM}^{(h_{\max})}_{\hat{l}_k-L}+\bR_{\hat{l}_k-L}^{(h_{\max})}), & L+1\leq\hat{l}_k\leq 2L,
\end{cases}
\ee
where $h_{\max}$ is the maximum number of EM iterations and
\be
\widehat{l}_k=\argmax_{l=1,\ldots,2L}q_k^{(h_{\max})}(l).
\ee
When the proposed decision schemes decide for $H_{1,i}$, $i=1,2,3$, the classification rule is that associated
with this hypothesis; on the contrary, data are partitioned according to the classification results under $H_0$.

\section{Numerical Examples and Discussion}
\label{Sec:performance}
In this section, we investigate the classification and detection performances of the proposed
architectures resorting to
standard Monte Carlo counting techniques.
Specifically, let us consider $N = 8$ spatial channels and two operating scenarios
differing in the number of targets, $L$, and $K$.
As for the clutter covariance components, we use $\bM_l = \sigma_{c,l}^2 \bM_c$, $l=1, \ldots, L$,
where $\sigma_{c,l}^2>0$ is the clutter power of the $l$th region
set according to the corresponding Clutter-to-Noise Ratio ($\mbox{CNR}_l$), while the $(i,j)$th element of $\bM_c$
is equal to $\rho^{|i-j|}$ with $\rho =0.9$. For simplicity, we assume the same
Signal-to-Interference-plus-Noise Ratio (SINR) value for all the synthetic targets.
More precisely, in the case of deterministic targets, it is
defined as $\mbox{SINR}= |\alpha_{k,l}|^2 \bv^\dag \bSigma_{l}^{-1} \bv$, whereas the
SINR for fluctuating targets is $\mbox{SINR} = \sigma_{k,l}^2 \small/(\sigma_{c,l}^2+\sigma_n^2)$,
where $\bSigma_{l}= \bM_l+\sigma_n^2 \bI$ with $\sigma_n^2=1$ the thermal
noise power, $\alpha_{k,l}$ and $\sigma_{k,l}^2$
are the amplitude and power associated with a target in the $k$th range bin of
the $l$th region,
the nominal steering vector is computed assuming that the AoAs are zero in all scenarios.
The penalty term in \eqref{eqn:E-step} is set
as $(N^2+k_{1,i}s)(1+\rho)/2$ with $\rho = 3$, $k_{1,1} = 2$, $k_{1,2} = 1$,
and $k_{1,3} = N$, that is borrowed from the
generalized information criterion with the parameter equal to $3$ \cite{Stoica1}.

\subsection{Operating scenario with two clutter regions}
\label{first-scenario}
Let us firstly focus on the scenario comprising two clutter regions of  $K_1 = K_2 = 32$ range bins. Each
region is characterized by $\bM_1$ and $\bM_2$ with $\mbox{CNR}_1 =20$ dB, $\mbox{CNR}_2 = 30$ dB.
Two targets appear in the $15$th range bin of the first region and in the $6$th range bin of the second
region. This operating scenario yields the following classes:
\begin{itemize}
  \item class 1: the generic vector of the first region does not contain any target component;
  \item class 2: the generic vector of the second region does not contain any target component;
  \item class 3: the generic vector of the first region contains  target components;
  \item class 4: the generic vector of the second region contains  target components.
\end{itemize}

As for the parameters initialization of the EM iterations under $H_0$ and $H_{1,i}$, $i = 1,2,3$,
we set $p_{l} = 1/L_c$, $l=1,\ldots,L_c$;
the initial value of $\bM_l$, namely $\widehat{\bM}_{l}^{(0)}$, is generated in the same way
as in Section \Rmnum{4}.A of \cite{clutterClust}.
A possible choice for $\widehat{\alpha}_{s,k}^{(0)}$, $s = 0,1$, $k=1,\ldots,K$, under $H_{1,1}$ is
\be
\widehat{\alpha}_{s,k}^{(0)}=
\begin{cases}
0, & s = 0,
\\
\dmax_{l=1,\ldots,L}
{\left(\ds\frac{\bv^\dag \widehat{\bM}_{l}^{(0)}\bz_k}{\bv^\dag \widehat{\bM}_{l}^{(0)}\bv}\right)}, & s = 1.
\end{cases}
\ee
Moreover, the initialization of $\sigma_{s,k}^2$ under $H_{1,2}$
is given by $(\widehat{\sigma}_{0,k}^2)^{(0)}=0$ and
$(\widehat{\sigma}_{1,k}^2)^{(0)}=|\bz_k^\dag \bv|^2, k=1,\ldots,K$. The initial value of $\bR_l$ under $H_{1,3}$ is grounded on a heuristic way:
1) compute $|\bz_k|^2$ for each range bin;
2) sort these quantities in descending order, namely $|\bz_{i_1}|^2 \geq |\bz_{i_2}|^2 \geq \ldots \geq |\bz_{i_K}|^2$,
where $i_k$ is the ordered index; 3) $\bR_l^{(0)}=\bz_{i_l} \bz_{i_l}^\dag, l=1,\ldots,L$.

\begin{figure}[tbp]
    \centering
    \includegraphics[width=.45\textwidth,height=3 cm]{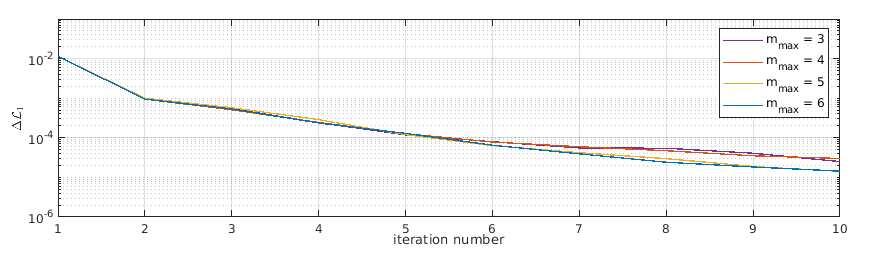}
    \caption{{$\Delta\cL_1(h)$ versus $h$ (number of iterations of the
    EM procedure) for different values of $m_{\max}$ (two targets, two clutter regions).}}
    \label{pconvergence_inner}
\end{figure}

\begin{figure}[tbp]
    \centering
    \includegraphics[width=.4\textwidth,height=5.5 cm]{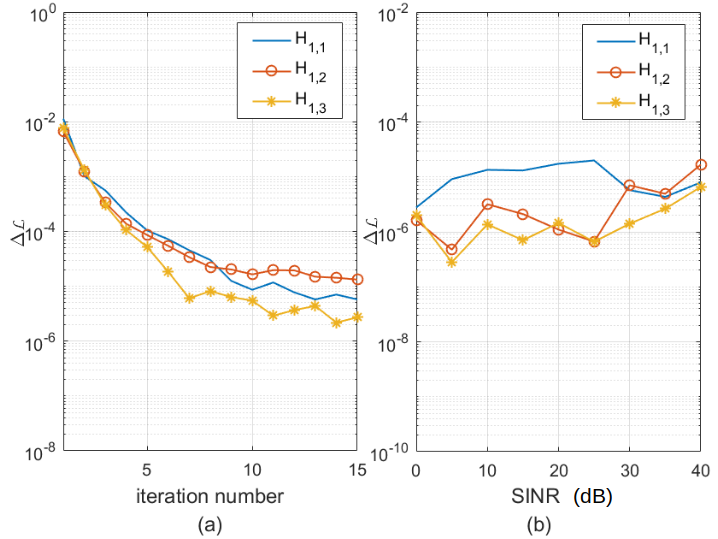}
    \caption{{The convergence performance of the EM procedure:
    (a) $\Delta\cL_i(h)$, $i=1,2,3$, versus $h$ (number of iterations of the
    EM procedure) for $\mbox{SINR} = 30$ dB;
    (b) $\Delta\cL_i(15)$, $i=1,2,3$, versus SINR (two targets, two clutter regions). }}
    \label{p1}
\end{figure}

Under the above assumptions, we first analyze the convergence performance of the
EM algorithm under the three target-plus-noise hypotheses.
To this end, let us define $\Delta \cL_i(h)= |[\cL(\bZ; \widehat{\cP}_{1,i}^{(h)})-\cL(\bZ; \widehat{\cP}_{1,i}^{(h-1)})]
/\cL(\bZ; \widehat{\cP}_{1,i}^{(h)})|$, $i=1,2,3$.
Then, the inner cyclic estimation
procedure under $H_{1,1}$ terminates when the convergence
criterion $\Omega(m) = {\| \widehat{\bM}_l^{(h-1),(m)} - \widehat{\bM}_l^{(h-1),(m-1)} \|}/{\| \widehat{\bM}_l^{(h-1),(m-1)} \|}
+{|\widehat{\alpha}_{1,k}^{(h-1),(m)}-\widehat{\alpha}_{1,k}^{(h-1),(m-1)}|}
/{|\widehat{\alpha}_{1,k}^{(h-1),(m-1)}|} < \delta$
is satisfied with $\delta = 10^{-4}$ or when $m = m_{\max}$. In the ensuing analysis,
we set $m_{\max}=5$ that ensures a good compromise between convergence and computational load.
In fact, Fig. \ref{pconvergence_inner}, where we plot the mean
curves (over $1000$ independent trials) of $\Delta\cL_1$ (namely, under $H_{1,1}$)
versus $h$ for different values of $m_{\max}$ and assuming $\mbox{SINR}=30$ dB,
confirms that $m_{\max}=5$ returns a relative variation of $\Delta\cL$
less than $10^{-4}$ when $h\geq 6$.
In Fig. \ref{p1}(a), the mean values of
$\Delta \cL_i(h)$, $i=1,2,3$,
are plotted resorting to $1000$ independent runs for $\mbox{SINR} = 30$ dB.
The convergence curves show that the log-likelihood variations
are roughly lower than $10^{-5}$ when at least 15 iterations are used
for the considered parameters.
The mean values of $\Delta \cL_i(15)$, $i=1,2,3$, versus SINR
are plotted in Fig. \ref{p1}(b) to verify that the variations maintain low values given $h_{\max}=15$.
Therefore, in the next illustrative examples, we set the maximum number of iteration $h_{\max}=15$.

\begin{figure*}[htb]
    \centering
    \subfigure[ ]{
    \begin{minipage}{0.3\linewidth}
    \centering
    \includegraphics[width=2in]{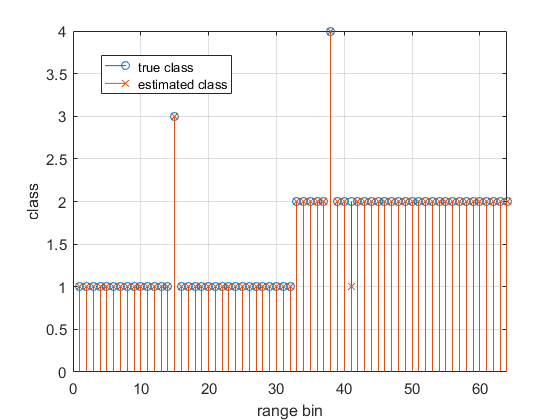}
    \end{minipage}
    }
    \subfigure[ ]{
    \begin{minipage}{0.3\linewidth}
    \centering
    \includegraphics[width=2in]{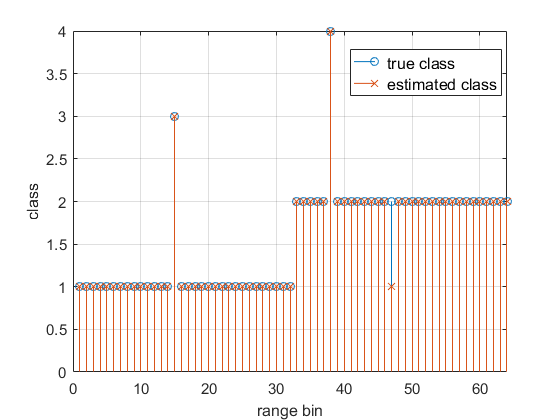}
    \end{minipage}
    }
    \subfigure[ ]{
    \begin{minipage}{0.3\linewidth}
    \centering
    \includegraphics[width=2in]{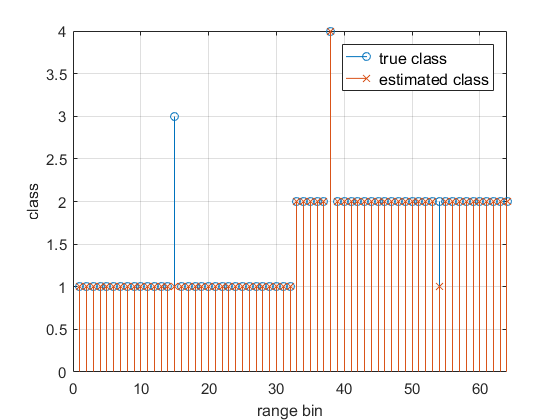}
    \end{minipage}}

    \subfigure[ ]{
    \begin{minipage}{0.3\linewidth}
    \centering
    \includegraphics[width=2in]{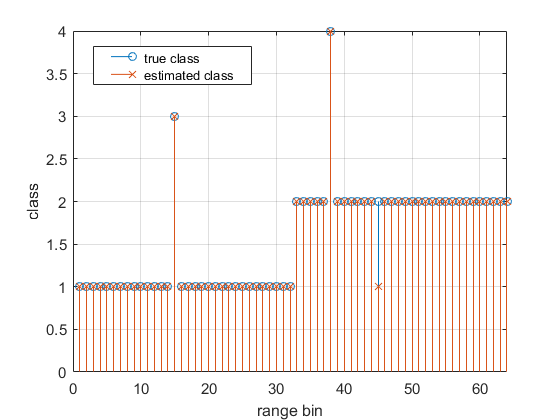}
    \end{minipage}
    }
    \subfigure[ ]{
    \begin{minipage}{0.3\linewidth}
    \centering
    \includegraphics[width=2in]{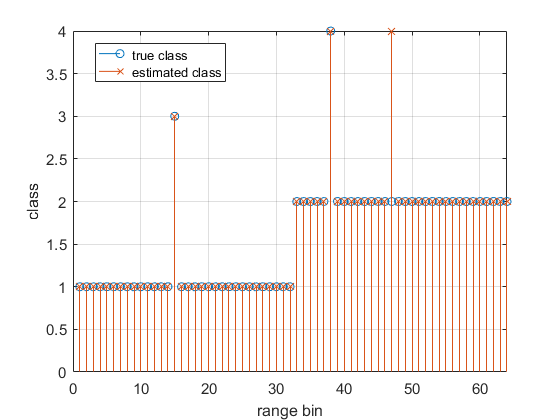}
    \end{minipage}
    }
    \subfigure[ ]{
    \begin{minipage}{0.3\linewidth}
    \centering
    \includegraphics[width=2in]{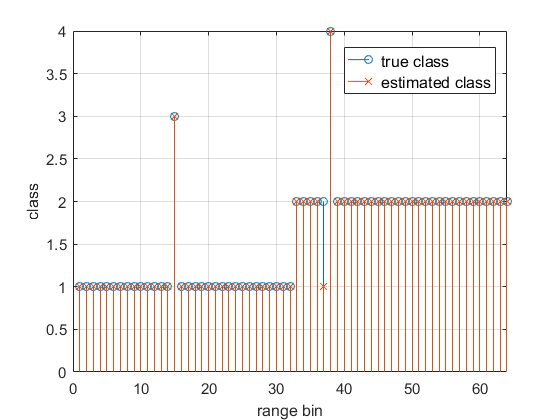}
    \end{minipage}
    }
    \subfigure[ ]{
    \begin{minipage}{0.3\linewidth}
    \centering
    \includegraphics[width=2in]{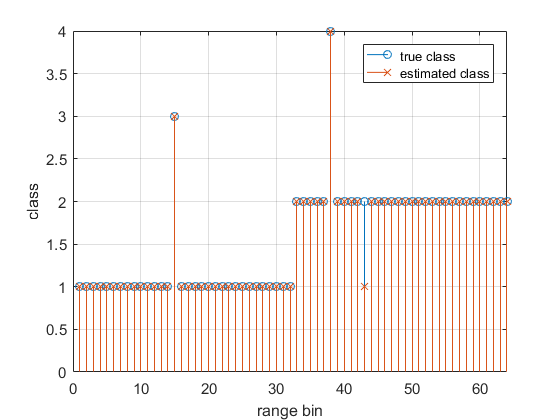}
    \end{minipage}
    }
    \subfigure[ ]{
    \begin{minipage}{0.3\linewidth}
    \centering
    \includegraphics[width=2in]{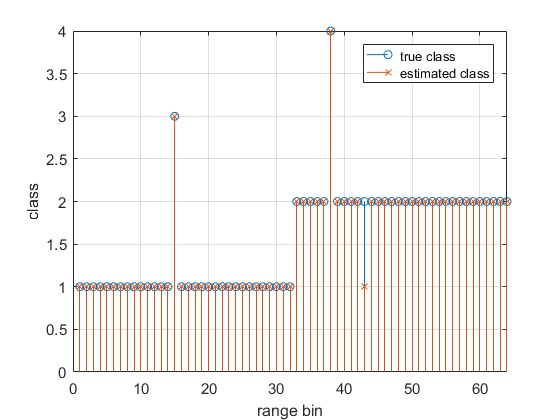}
    \end{minipage}
    }
    \subfigure[ ]{
    \begin{minipage}{0.3\linewidth}
    \centering
    \includegraphics[width=2in]{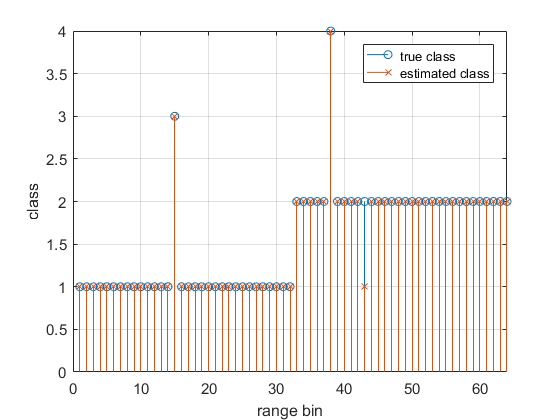}
    \end{minipage}
    }
    \caption{Classification snapshot for different SINRs (two targets, two clutter regions):
    (a) $\mbox{SINR} = 15$ dB under $H_{1,1}$; (b) $\mbox{SINR} = 15$ dB under $H_{1,2}$; (c) $\mbox{SINR} = 15$ dB under $H_{1,3}$;
    (d) $\mbox{SINR} = 25$ dB under $H_{1,1}$; (e) $\mbox{SINR} = 25$ dB under $H_{1,2}$; (f) $\mbox{SINR} = 25$ dB under $H_{1,3}$;
    (g) $\mbox{SINR} = 35$ dB under $H_{1,1}$; (h) $\mbox{SINR} = 35$ dB under $H_{1,2}$; (i) $\mbox{SINR} = 35$ dB under $H_{1,3}$.}
    \label{p2}

\end{figure*}

\begin{figure}[htb]
    \centering
    \includegraphics[width=.5\textwidth,height=4 cm]{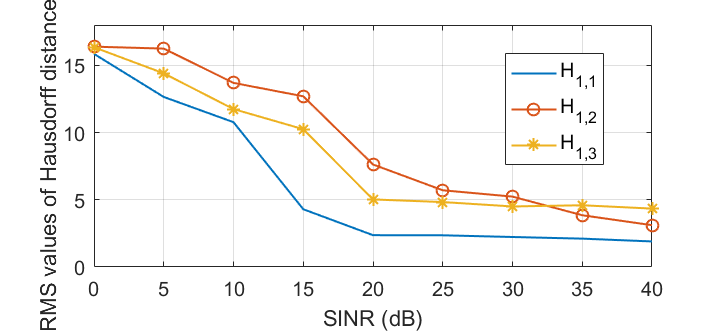}
    \caption{Hausdorff distance versus SINR under $H_{1,i}$, $i=1,2,3$ (two targets, two clutter regions).}
    \label{pHausdorff1}
\end{figure}

In Fig. \ref{p2}, we plot a snapshot (that is, one Monte Carlo trial)
of the classification results under each alternative hypothesis
for $\mbox{SINR}=15,25,35$ dB, leaving unaltered the
other parameters. A qualitative inspection of the results highlights that the
three models lead to almost equivalent classification performances.
In Table \ref{T1}, we show the Root Mean Square Classification Error (RMSCE) values with respect
to the covariance matrix class (its definition can be found in \cite{clutterClust} and is omitted here
for brevity) that
confirms the above conclusion with the estimation procedure under $H_{1,1}$ ensuring slightly better
performance than the other procedures for low SINR values. On the other hand,
for high SINR values the procedures under $H_{1,2}$ and $H_{1,3}$ overcome that under $H_{1,1}$.
In order to measure the error in target
position estimation, we estimate the Root Mean Square (RMS) values of the Hausdorff metric \cite{4567674,ECCMYan}
over $1000$ trials in Fig. \ref{pHausdorff1} between $\widehat{\bor}$ and $\bor$.
Precisely, $\widehat{\bor}$ is a vector of size $K$
that contains 0 except for the range bins classified as target,
whereas $\bor$ is an analogous vector containing the true range bin positions of the targets.
The figure points out that the classifier under $H_{1,1}$ has lower RMS values than the other two
procedures that suffer the existence of more ghosts. Since the effects of
the clutter power levels on the classification performance have been assessed in \cite{clutterClust},
we omit this part of the analysis.

\begin{figure}[tbp]
    \centering
    \includegraphics[width=.43\textwidth, height=5.3cm]{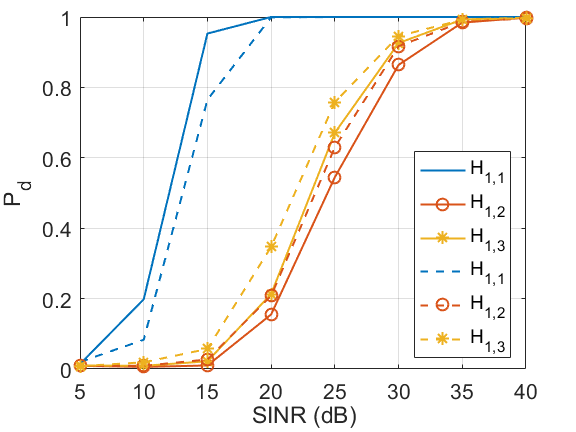}
    \caption{$P_d$ of the detectors in (\ref{eqn:LRT_1}) (dashed line)
    and (\ref{eqn:LRT_LVM}) (solid line) versus SINR under $H_{1,i}$, $i=1,2,3$, assuming $P_{fa}=10^{-2}$
    (two targets, two clutter regions).}
    \label{p3}
\end{figure}

\begin{figure*}[htb]
    \centering
    \subfigure[ ]{
    \begin{minipage}{0.3\linewidth}
    \centering
    \includegraphics[width=2in]{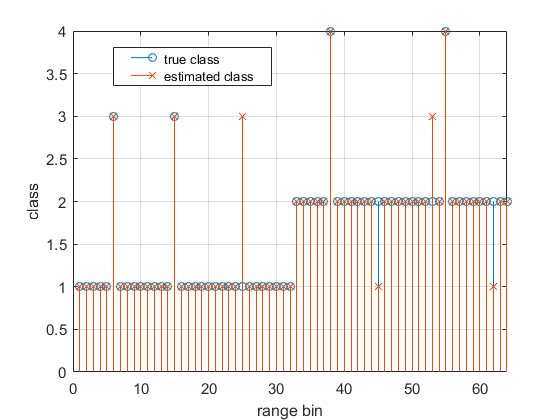}
    \end{minipage}
    }
    \subfigure[ ]{
    \begin{minipage}{0.3\linewidth}
    \centering
    \includegraphics[width=2in]{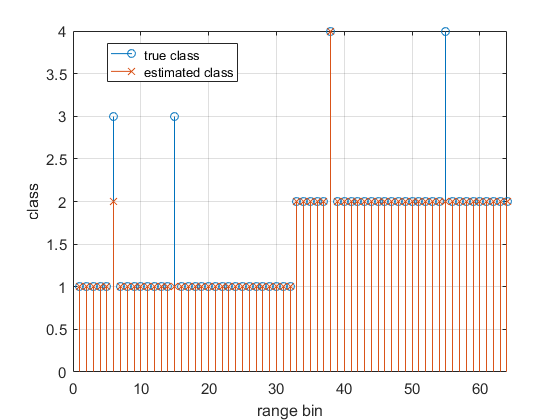}
    \end{minipage}
    }
    \subfigure[ ]{
    \begin{minipage}{0.3\linewidth}
    \centering
    \includegraphics[width=2in]{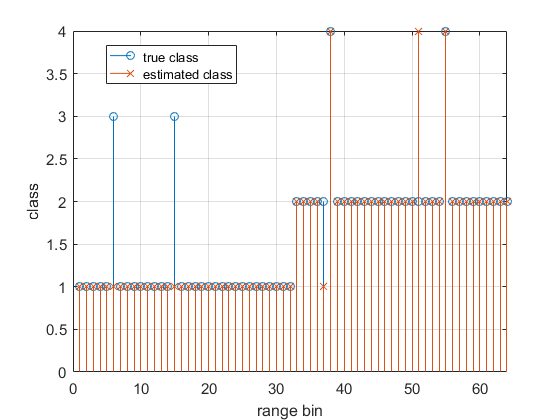}
    \end{minipage}}

    \subfigure[ ]{
    \begin{minipage}{0.3\linewidth}
    \centering
    \includegraphics[width=2in]{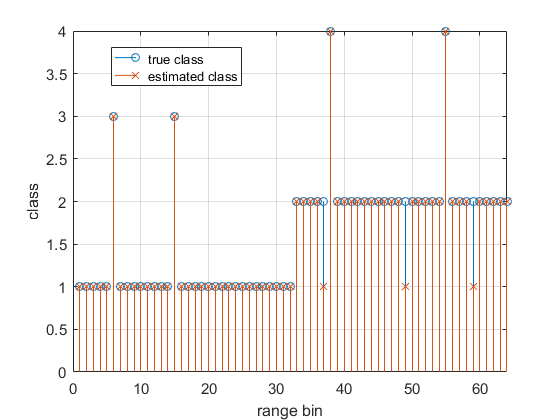}
    \end{minipage}
    }
    \subfigure[ ]{
    \begin{minipage}{0.3\linewidth}
    \centering
    \includegraphics[width=2in]{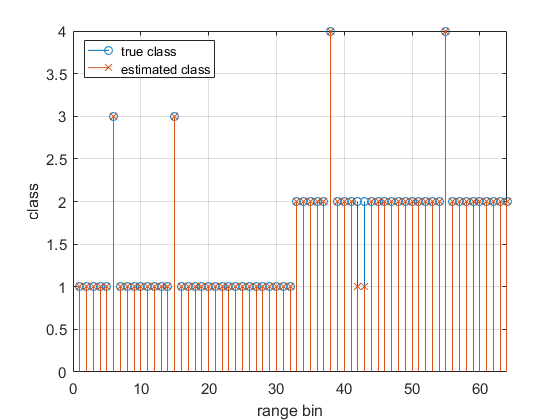}
    \end{minipage}
    }
    \subfigure[ ]{
    \begin{minipage}{0.3\linewidth}
    \centering
    \includegraphics[width=2in]{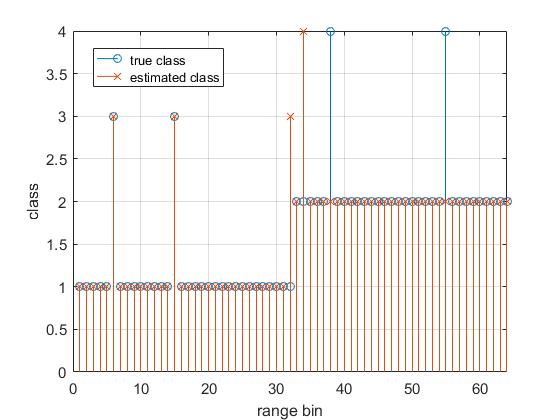}
    \end{minipage}
    }
    \subfigure[ ]{
    \begin{minipage}{0.3\linewidth}
    \centering
    \includegraphics[width=2in]{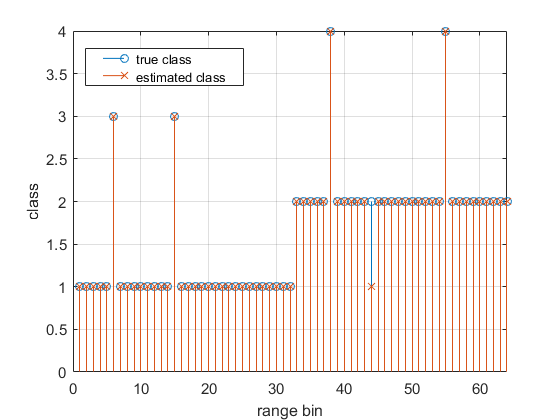}
    \end{minipage}
    }
    \subfigure[ ]{
    \begin{minipage}{0.3\linewidth}
    \centering
    \includegraphics[width=2in]{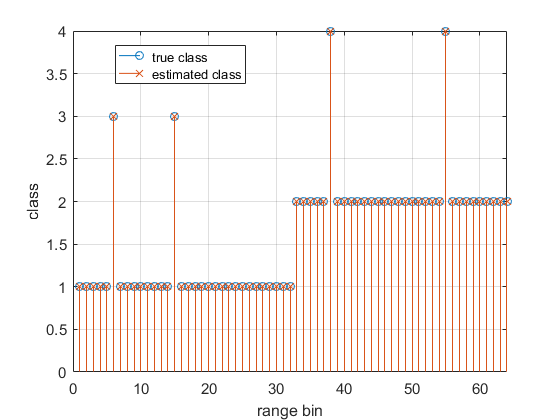}
    \end{minipage}
    }
    \subfigure[ ]{
    \begin{minipage}{0.3\linewidth}
    \centering
    \includegraphics[width=2in]{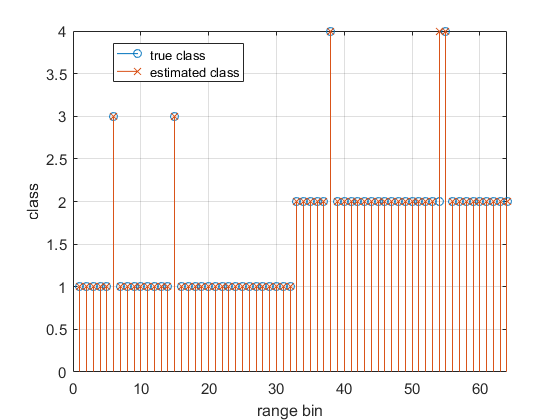}
    \end{minipage}
    }
    \caption{Classification snapshot for different SINRs (four targets, two clutter regions):
    (a) $\mbox{SINR} = 15$ dB under $H_{1,1}$; (b) $\mbox{SINR} = 15$ dB under $H_{1,2}$; (c) $\mbox{SINR} = 15$ dB under $H_{1,3}$;
    (d) $\mbox{SINR} = 25$ dB under $H_{1,1}$; (e) $\mbox{SINR} = 25$ dB under $H_{1,2}$; (f) $\mbox{SINR} = 25$ dB under $H_{1,3}$;
    (g) $\mbox{SINR} = 35$ dB under $H_{1,1}$; (h) $\mbox{SINR} = 35$ dB under $H_{1,2}$; (i) $\mbox{SINR} = 35$ dB under $H_{1,3}$.}
    \label{p4}
\end{figure*}

\begin{figure}[htb]
    \centering
    \includegraphics[width=.5\textwidth,height=4cm]{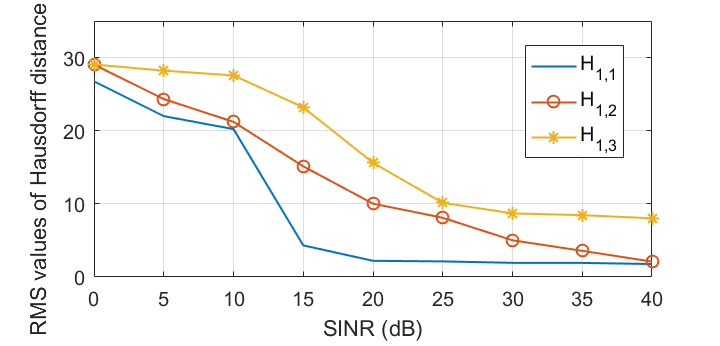}
    \caption{Hausdorff distance versus SINR under $H_{1,i}$, $i=1,2,3$ (four targets, two clutter regions).}
    \label{pHausdorff2}
\end{figure}

\begin{figure}[htb]
    \centering
    \includegraphics[width=.43\textwidth,height=5cm]{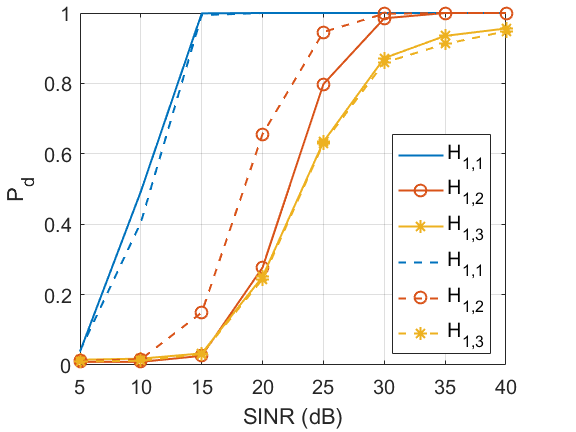}
    \caption{{$P_d$ of the detectors (\ref{eqn:LRT_1}) (dashed line)
    and (\ref{eqn:LRT_LVM}) (solid line) versus SINR under $H_{1,i}$, $i=1,2,3$, assuming $P_{fa}=10^{-2}$
    (four targets, two clutter regions).}}
    \label{p5}
\end{figure}

\begin{table}
  \centering
  \caption{RMSCE values under $H_{1,i}$, $i=1,2,3$, for different SINRs (two targets, two clutter regions)}
  \begin{tabular}{|c|c|c|c| p{2.5 cm}}
  \hline
    & $H_{1,1}$ & $H_{1,2}$ & $H_{1,3}$\\
  \hline
  SINR = 15 dB & 4.1490 & 4.3380 & 4.3906\\
  \hline
  SINR = 25 dB & 3.7114 & 3.0842 & 2.9630\\
  \hline
  SINR = 35 dB & 3.4347 & 2.7841 & 2.6029 \\ \hline

  \end{tabular}
  \label{T1}
\end{table}
In Fig. \ref{p3}, we evaluate the Probability
of Detection ($P_d$) returned by detectors \eqref{eqn:LRT_1} and \eqref{eqn:LRT_LVM}
as a function of the SINR under each hypothesis.
Specifically, the detection thresholds
are set over $100/P_{fa}$ independent runs with $P_{fa}=10^{-2}$,
whereas the $P_d$ is computed exploiting 1000 independent trials.
The $P_d$ values associated with \eqref{eqn:LRT_1} are slightly lower than those
of the detector \eqref{eqn:LRT_LVM} under $H_{1,1}$. The opposite behavior can be observed under
$H_{1,2}$ and $H_{1,3}$. Moreover, both detectors for $H_{1,3}$ return slightly better performance
than the detectors for $H_{1,2}$ in the medium/high SINR region.
The last example in this subsection assumes the presence of additional targets occupying
the $6$th and $55$th range bins. The classification results are reported in Fig. \ref{p4}
where for $\mbox{SINR} = 15$ dB, the classifier under $H_{1,1}$ is capable of correctly identifying the
range bins with target components whereas the other classifiers miss two targets.
When the SINR increases to $25$ dB the classifiers under $H_{1,1}$ and $H_{1,2}$ ensure a reliable classification,
whereas that for $H_{1,3}$ still misses targets. Finally, when $\mbox{SINR} = 35$ dB, the performance
of all classifiers becomes excellent.
As for the Hausdorff metric, in Fig. \ref{pHausdorff2}, the classifier for $H_{1,1}$ provides
the lowest RMS errors, while that for $H_{1,3}$ is the worst and converges to a larger
constant with respect to Fig. \ref{pHausdorff1}. In Table \ref{T2},
the RMSCE values with respect to the covariance class are larger than those
in Table \ref{T1} due to the classification loss.
Finally, the $P_d$ curves are shown in Fig. \ref{p5}. It turns out that detectors for $H_{1,1}$ and $H_{1,2}$
increase their detection performance with respect to the previous case with two targets, whereas
detectors for $H_{1,3}$ experience a performance degradation due to the incorrect classification results.

\begin{table}
  \centering
  \caption{RMSCE values under $H_{1,r}, r=1,2,3$, for different SINRs (four targets, two clutter regions)}
  \begin{tabular}{|c|c|c|c| p{2.5 cm}}
  \hline
    & $H_{1,1}$ & $H_{1,2}$ & $H_{1,3}$\\
  \hline
   SINR = 15 dB & 5.9711 & 6.7064 & 7.4718\\
  \hline
  SINR = 25 dB & 5.4401 & 4.2819 & 7.1379\\
  \hline
  SINR = 35 dB & 4.9620 & 3.8834 & 6.9627 \\ \hline

  \end{tabular}
  \label{T2}
\end{table}

\subsection{Operating scenario with three clutter regions}
In this subsection, we consider three clutter regions of
$K_1 = K_2 = K_3 = 32$ range bins. The clutter power of these regions is set
according to $\mbox{CNR}_1=20$ dB, $\mbox{CNR}_2=30$ dB, and $\mbox{CNR}_3=40$ dB.
Moreover, we include four targets in the $16$th, $36$th, $75$th, and $85$th
range bin.
The values of the other parameters and initialization are the same as those in the Figs. \ref{p2}-\ref{p3}.
In this scenario, the number of considered classes becomes $6$, namely,
\begin{itemize}
  \item class 1: the generic vector of the first region does not contain any target component;
  \item class 2: the generic vector of the second region does not contain any target component;
  \item class 3: the generic vector of the third region does not contain any target component;
  \item class 4: the generic vector of the first region contains target components;
  \item class 5: the generic vector of the second region contains target components;
  \item class 6: the generic vector of the third region contains target components.
\end{itemize}
The convergence behavior is almost the same as in the previous subsection and is not reported here
for brevity. Thus, also in this case we set $m_{\max}=5$ and $h_{\max}=15$.

\begin{figure*}[htb]
    \centering
    \subfigure[ ]{
    \begin{minipage}{0.3\linewidth}
    \centering
    \includegraphics[width=2in]{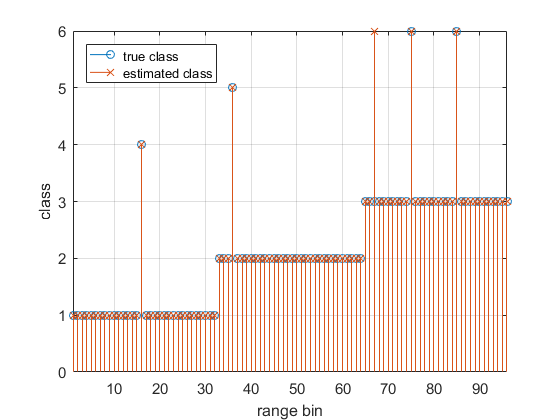}
    \end{minipage}
    }
    \subfigure[ ]{
    \begin{minipage}{0.3\linewidth}
    \centering
    \includegraphics[width=2in]{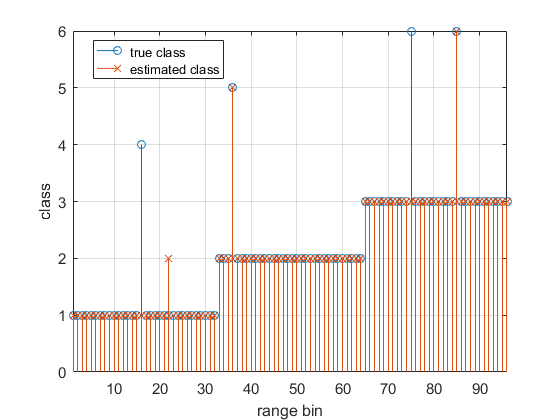}
    \end{minipage}
    }
    \subfigure[ ]{
    \begin{minipage}{0.3\linewidth}
    \centering
    \includegraphics[width=2in]{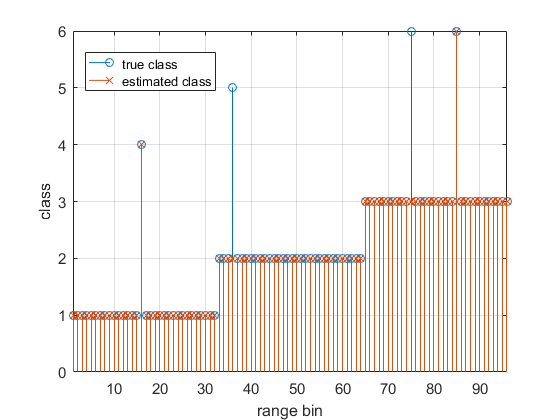}
    \end{minipage}}

    \subfigure[ ]{
    \begin{minipage}{0.3\linewidth}
    \centering
    \includegraphics[width=2in]{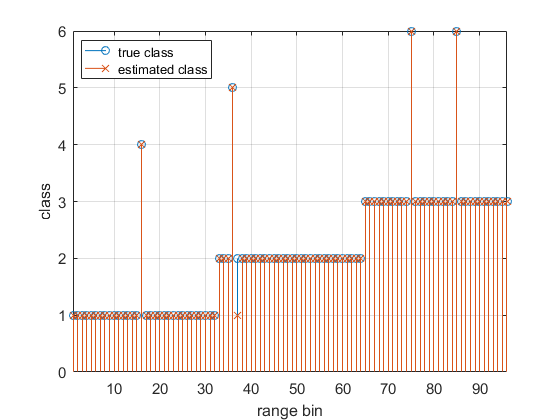}
    \end{minipage}
    }
    \subfigure[ ]{
    \begin{minipage}{0.3\linewidth}
    \centering
    \includegraphics[width=2in]{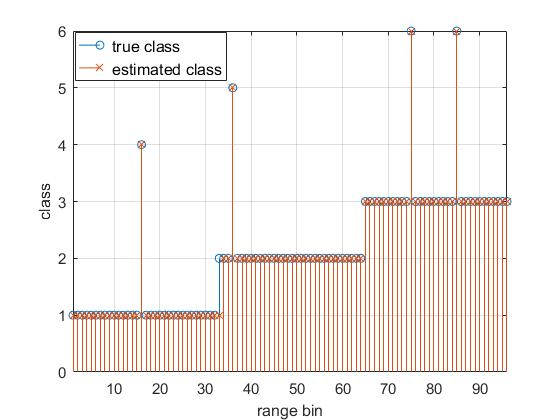}
    \end{minipage}
    }
    \subfigure[ ]{
    \begin{minipage}{0.3\linewidth}
    \centering
    \includegraphics[width=2in]{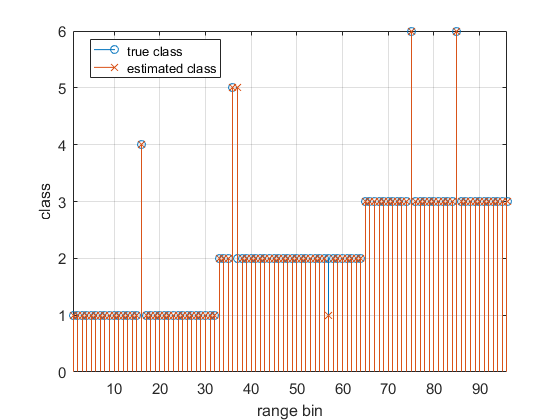}
    \end{minipage}
    }
    \subfigure[ ]{
    \begin{minipage}{0.3\linewidth}
    \centering
    \includegraphics[width=2in]{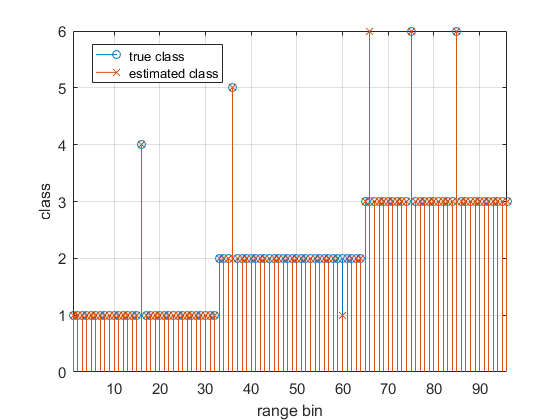}
    \end{minipage}
    }
    \subfigure[ ]{
    \begin{minipage}{0.3\linewidth}
    \centering
    \includegraphics[width=2in]{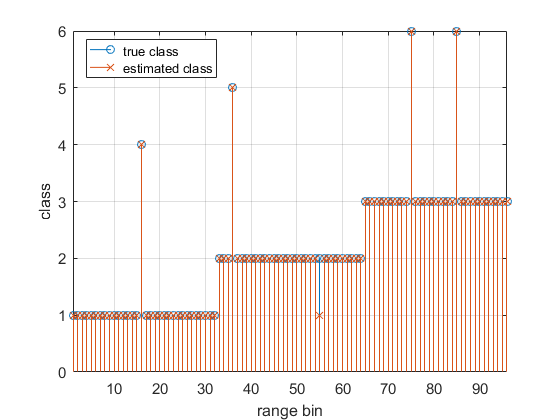}
    \end{minipage}
    }
    \subfigure[ ]{
    \begin{minipage}{0.3\linewidth}
    \centering
    \includegraphics[width=2in]{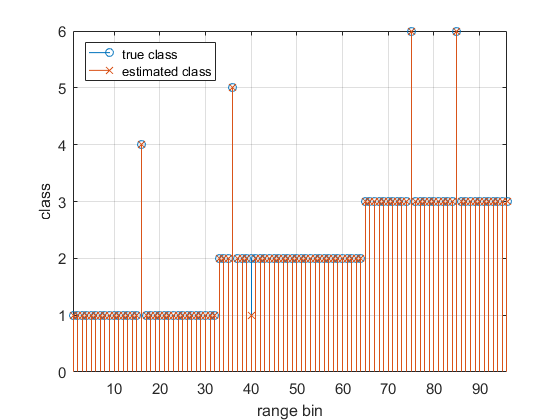}
    \end{minipage}
    }
    \caption{Classification snapshot for different SINRs (three clutter regions):
    (a) $\mbox{SINR} = 15$ dB under $H_{1,1}$; (b) $\mbox{SINR} = 15$ dB under $H_{1,2}$; (c) $\mbox{SINR} = 15$ dB under $H_{1,3}$;
    (d) $\mbox{SINR} = 25$ dB under $H_{1,1}$; (e) $\mbox{SINR} = 25$ dB under $H_{1,2}$; (f) $\mbox{SINR} = 25$ dB under $H_{1,3}$;
    (g) $\mbox{SINR} = 35$ dB under $H_{1,1}$; (h) $\mbox{SINR} = 35$ dB under $H_{1,2}$; (i) $\mbox{SINR} = 35$ dB under $H_{1,3}$.}
    \label{p9}

\end{figure*}

\begin{figure}[htb]
    \centering
    \includegraphics[width=.5\textwidth,height=4cm]{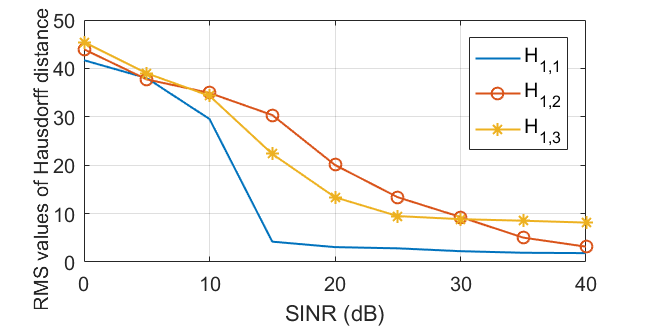}
    \caption{Hausdorff distance versus SINR under $H_{1,i}$, $i=1,2,3$ (three clutter regions).}
    \label{pHausdorff3}
\end{figure}

\begin{figure}[htb]
    \centering
    \includegraphics[width=.43\textwidth,height=5cm]{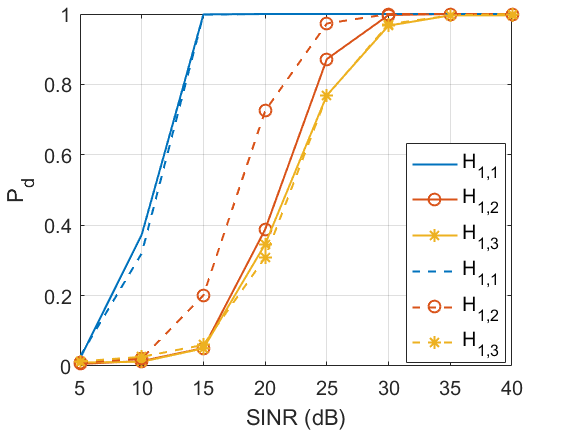}
    \caption{$P_d$ of the detectors (\ref{eqn:LRT_1}) (dashed line)
    and (\ref{eqn:LRT_LVM}) (solid line) versus SINR
    under $H_{1,i}$, $i=1,2,3$, assuming $P_{fa}=10^{-2}$ (three clutter regions).}
    \label{p10}
\end{figure}

A qualitative assessment of the classification capabilities can be obtained through Fig. \ref{p9}, where
the classification procedure under $H_{1,1}$ turns out to be more robust than the other procedures
for $\mbox{SINR}=15$ dB. Moreover, as in the previous case, the higher the SINR, the more reliable the classification results.
The RMSCE values with respect to the clutter class
are reported in Table \ref{T4} and confirm what observed in Fig. \ref{p9}
from another perspective.
The Hausdorff curves for this scenario are contained
in Fig. \ref{pHausdorff3} and exhibit a similar trend as in Fig. \ref{pHausdorff1}.
As a matter of fact, also in this case, the classification procedure under $H_{1,1}$
returns the best performance in terms of target position estimation.
Finally, the corresponding $P_d$ curves are confined to Fig. \ref{p10} where
the highest $P_d$ values are returned by the detectors for $H_{1,1}$. As in Fig. \ref{p5},
also in this case, the curves of decision schemes for $H_{1,3}$ come after those of the other
architectures.

\begin{table}
  \centering
  \caption{RMSCE values under $H_{1,i}$, $i=1,2,3$, for different SINRs (three clutter regions)}
  \begin{tabular}{|c|c|c|c| p{2.5 cm}}
  \hline
    & $H_{1,1}$ & $H_{1,2}$ & $H_{1,3}$\\
  \hline
   SINR = 15 dB & 4.2706 & 4.5399 & 4.9164\\
  \hline
  SINR = 25 dB & 3.6266 & 3.0715 & 4.1104\\
  \hline
  SINR = 35 dB & 3.4316 & 2.7658 & 2.9159\\ \hline

  \end{tabular}
  \label{T4}
\end{table}

\section{Conclusions}
\label{Sec:conclusions}
In this paper, we have devised detection architectures dealing with multiple point-like targets
in heterogeneous scenarios. At the design stage, neither the number of targets nor their positions
have been assumed known as well as the clutter regions within the data window.
In this context, we have devised three estimation procedures based upon different signal models.
Their common denominator is the EM algorithm that has been suitably modified and/or approximated
in order to come up with closed-form updates for the parameter estimates.
Then, such estimates have been used to implement two decision schemes relying on the LRT.
The classification and detection capabilities of the proposed architectures
have been assessed over synthetic data simulating different operating scenarios
with an increasing complexity in terms of the number of clutter regions and targets.
This analysis has shown that the proposed architectures can provide a rather likely picture of the entire operating
scenarios making the radar system aware of the surrounding environment.

Future research tracks might encompass the design of cognitive schemes that account for more specific
covariance structures or, more importantly, that consider the joint presence of point-like
as well as range-spread targets.

\section{ACKNOWLEDGMENTS}
This work was supported by the National Natural Science Foundation of China under Grant No. 61571434.

\bibliographystyle{IEEEtran}
\bibliography{group_bib}

\end{document}